\documentclass[prd, onecolumn, floatfix, superscriptaddress]{revtex4}
\usepackage[utf8]{inputenc}
\pdfoutput=1
 
\usepackage{graphicx}
\usepackage{epsfig}
\usepackage{bm}
\usepackage{amssymb}
\usepackage{float}
\usepackage{amsmath}
\usepackage{dcolumn}
\usepackage{cancel}
\usepackage[colorlinks]{hyperref}
\usepackage[usenames, dvipsnames]{color}

\hypersetup{
     breaklinks=true, 
    pdfstartview={FitH},  
    colorlinks=true, 
    linkcolor=blue,  
    citecolor=red,  
    filecolor=magenta,  
    urlcolor=blue, 
    anchorcolor=green,  
    linktocpage=true
}

\def\doi{http://doi.org}

\begin{document}

\title{Fermion localization in a extra-dimensional $f(Q,\mathcal{T})$ gravity 
with cuscuton dynamics}

\author{A. R. P. Moreira}
\affiliation{Research Center for Quantum Physics, Huzhou University, Huzhou, 
313000, P. R. China.}
\affiliation{Secretaria da Educaç\~{a}o do Cear\'{a} (SEDUC), Coordenadoria 
Regional de Desenvolvimento da Educaç\~{a}o (CREDE 9),  Horizonte, Cear\'{a}, 
62880-384, Brazil.}
\author{Shi-Hai Dong}
\affiliation{Research Center for Quantum Physics, Huzhou University, Huzhou, 
313000, P. R. China.}
\affiliation{Centro de Investigaci\'{o}n en Computaci\'{o}n, Instituto 
Polit\'{e}cnico Nacional, UPALM, CDMX 07700, Mexico.}

\author{Emmanuel N. Saridakis}
\affiliation{National Observatory of Athens, Lofos Nymfon, 11852 Athens, Greece}
\affiliation{CAS Key Laboratory for Researches in Galaxies and Cosmology, 
School 
of Astronomy and Space Science, University of Science and Technology of China, 
Hefei, Anhui 230026, China}
\affiliation{Departamento de Matem\'{a}ticas, Universidad Cat\'{o}lica del 
Norte, Avda. Angamos 0610, Casilla 1280 Antofagasta, Chile}

\begin{abstract}
We investigate the fermion localization in extra-dimensional $f(Q,\mathcal{T})$ 
gravity with cuscuton dynamics. This modified gravity theory is based on the  
nonmetricity scalar $Q$ and on the trace of the energy-momentum tensor 
$\mathcal{T}$. Using the first-order formalism  we construct the complete brane 
system for three well-known superpotentials, namely the Sine-Gordon, the 
polynomial and the  linear one. We show that the addition of the cuscuton term  
provides significant modifications to the structure of the brane. In particular, 
  the scalar field solutions have the form of a kink-like structure, and the 
energy density is well localized, depending on both the modified-gravity and the 
cuscuton parameters. Furthermore,  applying probabilistic measures, for the 
location of the fermions for a minimal Yukawa-type coupling we arrive
at a Schrödinger-like equation allowing for a normalizable massless mode. 
Our solutions  indicate strong brane localization only for 
left-chirality  fermions.  Moreover,   the massive modes present 
solutions similar to free waves, which indicates that these fermions probably 
escape the brane. Finally, we find  that massive fermions have a greater 
sensitivity to gravitational changes in the core of the brane, where 
oscillations with more pronounced amplitudes are present. 

\end{abstract}
 
\maketitle

\section{Introduction}

Nearly a century ago, Kaluza and Klein published a remarkable work in which the 
hypothesis of extra dimension was considered as an attempt of combining gravity 
and electrodynamics in a single theory \cite{Kaluza:1921tu,Klein:1926tv}. 
Despite being an elegant hypothesis, extra dimensions suffer in the scientific 
community due to the lack of experimental evidence. Proposed in 1999, the 
so-called Randall-Sundrum (RS) model reopened the theoretical viability of extra 
dimensions \cite{rs,rs2}, aiming to solve the hierarchy problem, an important   
question in theoretical physics for many years. 

Higher-dimensional models, also known as braneworld, have been   investigated in 
the literature over the past two decades. In particular, thick branes have been 
a direct extension of the RS model, where a real scalar field is introduced to 
provide thickness and smooth warp factor for the brane 
\cite{Gremm1999,Bazeia2008,Dzhunushaliev2009,Charmousis2001,Arias2002,
Barcelo2003,Bazeia2004,CastilloFelisola2004, Dutra2014}. Another important 
feature of thick branes is the appearance of internal structure.

Solitons kinks-like, vortices, and monopoles are  topological structures that 
have played an important role in both high-energy and condensed matter physics, 
appearing in several models that exhibit spontaneous symmetry breaking as it 
occurs in the phenomena of superconductivity  
\cite{Abrikosov:1956sx,Davis:1988jp} and planar physics \cite{Bais:2002pb}. Some 
of these structures like the kinks are modeled by real scalar fields, which 
have, in general, a standard kinetic term and a potential 
\cite{Bazeia:1995en,Bazeia:2002xg}. Nevertheless, noncanonical models have also 
attracted attention. An example of noncanonical dynamics is the cuscuton theory 
proposed by Afshordi \cite{Afshordi:2006ad}. Such a theory is regarded as a 
field with infinite speed of propagation and has its origin when the 
Hamiltonian analysis in the homogeneous limit is performed. It has been shown 
that the cuscuton term does not represent an additional degree of freedom, being 
as an auxiliary field. Recently, special attention has been given to cuscuton 
dynamics in cosmology 
\cite{Afshordi:2007yx,HosseiniMansoori:2022xnq,Bartolo:2021wpt}, vortices 
\cite{Lima:2021ekz,Lima:2023str} and in the context of five-dimensional 
braneworlds 
\cite{Andrade:2018afh,Bazeia:2021jok,Bazeia:2021bwg,Rosa:2021myu,
Bazeia:2022sgb}.

In the thick brane scenarios,  besides modifying the scalar field dynamics, one 
can consider modified gravity theories \cite{CANTATA:2021asi}, such  as $f(R)$ 
and $f(R,\mathcal{T})$ gravities 
\cite{Afonso:2007gc, Bazeia:2014poa,Gu:2014ssa}, 
where $R$ represents the curvature scalar and $\mathcal{T}$ the trace of 
momentum-energy tensor. In addition to these theories, there are curvature-free 
theories based on non-Riemannian geometry as $f(T)$, $f(T,B)$ $f(T,\mathcal{T})$ 
and $f(Q)$ \cite{Yang2012,Menezes,tensorperturbations, ftnoncanonicalscalar, 
ftborninfeld,ftmimetic,Yang2017,Moreira:2021xfe,Moreira20211,Moreira:2021wkj, 
Moreira:2021uod,Belchior,Fu:2021rgu,Silva:2022pfd}, where $T$ is torsion scalar, 
$B$ the boundary term and $Q$ nonmetricity scalar 
\cite{Cai:2015emx,Heisenberg:2023lru}. Herein, $f(T)$, $f(T,B)$ and 
$f(T,\mathcal{T})$ gravities are modifications of the teleparallel equivalent of 
general relativity (TEGR) \cite{Cai:2011tc,ftenergyconditions, Iorio:2012cm,
Bahamonde2015,Ferraro2011us,Tamanini2012,Harko:2014sja,Bahamonde2016,
Franco2020, 
EscamillaRivera2019,Bahamonde2016a,Caruana2020,ElHanafy:2020pek,Pourbagher2020, 
Bahamonde2020a, Harko:2014aja,Wang:2023qfm}, and $f(Q)$ gravity is a 
modification of symmetric teleparallel equivalent of general relativity (STEGR) 
\cite{BeltranJimenez:2019esp,BeltranJimenez:2017tkd,Anagnostopoulos:2021ydo,
BeltranJimenez:2019tme, 
Bajardi:2020fxh,Anagnostopoulos:2022gej,Capozziello:2022tvvi,Capozziello:2022wgl
, Gadbail:2022jco, 
Wang:2024eai,Wu:2024vcr}.

The great success of braneworld constructions and modified theories of gravity 
has motivated many researchers to study braneworld scenarios within the modified 
gravity frame. In this work, we  are interested in studying five-dimensional 
thick branes in gravity $f(Q,\mathcal{T})$ coupled to a single scalar field with 
cuscuton dynamics, and thus going beyond the analysis of 
\cite{Moreira:2024unj,Belchior:2025vad}. In particular, we will apply the 
first-order formalism   
 to find analytical solutions for our brane system. Furthermore, we will 
investigate the fermion locations  using probabilistic 
measurements.

The outline of this work is the following:  In section \ref{s2} we
introduce the essential concepts of STERG and we obtain the equations of 
motions for $f(Q,\mathcal{T})$ gravity. In section \ref{s3} we construct the 
cuscuton brane  within $f(Q,\mathcal{T})$ gravity by employing 
the first-order formalism. In section \ref{s4} we verify whether the 
brane system is stable under small perturbations and we study the localization 
of massive and resonates modes for graviton. Our 
final discussion as well as future perspectives are presented in section 
\ref{s5}.

\section{modified symmetric teleparallel gravity}
\label{s2}

As starting point, let us  introduce the general concepts of symmetric 
teleparallel gravity and present the equation of motion for $f(Q,\mathcal{T})$ 
gravity. When we deal with theories based on Riemannian geometry 
the metricity condition $\nabla_M g_{NP}=0$ must be satisfied, where $g_{NP}$ 
represents the metric and $\nabla_M$ is the covariant derivative with the 
Levi-Civita $\Gamma^P\ _{MN}$ as the affine connection. The bulk coordinate 
indices are denoted by capital Latin index $M=0,\ldots,D-1$. Such condition is 
obeyed by GR and $f(R)$ gravity for instance. However, this relation is no 
longer satisfied in theories based on non-Riemannian geometry. Among these 
modified gravity models, symmetric teleparallel equivalent of general relativity 
(STEGR)  and its main feature is the presence of nonvanishing nonmetricity 
tensor \cite{Nester:1998mp}
\begin{equation}
Q_{MNP}=\nabla_M g_{NP}.   
\end{equation}
For this tensor, we define its independent traces  $Q_M=g^{NP}Q_{MNP}$ and 
$\widetilde{Q}_M=g^{NP}Q_{NMP}$. In addition, due to the nonmetricity tensor, a 
more general connection $\widetilde{\Gamma}^P\ _{MN}$ for STEGR must be defined 
as 
\begin{equation}
\widetilde{\Gamma}^P\ _{MN}=\Gamma^P\ _{MN}+L^P\ _{MN},    
\end{equation}
where $L^P\ _{MN}$ is known as the distortion tensor, which is written in  
nonmetricity tensor terms as  \cite{Nester:1998mp}
\begin{equation}
L^P\ _{MN}=\frac{1}{2}g^{PQ}(Q_{PMN}-Q_{MPN}-Q_{NPM}).    
\end{equation}

To construct a gravitational action for STEGR, we conveniently introduce a more 
general tensor that contains the nonmetricity, its independent traces, and 
distortion tensor. This tensor is  known as nonmetricity conjugate given by 
\cite{Nester:1998mp}
\begin{equation}
P^P\ _{MN}=-\frac{1}{2}L^P\ _{MN}+\frac{1}{4}(Q^P-\widetilde{Q}^P)g_{MN}-\frac{1}{8}(\delta^P_M Q_N+\delta^P_N Q_M).    
\end{equation}
Besides, its contraction with nonmetricity tensor  provides the nonmetricity 
scalar $Q=Q_{PMN}P^{PMN}$. 

The Ricci scalar is written as $R=Q+B$, where $B=\nabla_M(Q^M-\widetilde{Q}^M)$ 
is a boundary term. Such a result shows that STEGR is equivalent to GR since the 
boundary term vanishes when  integrated into the action. However,   for $f(Q)$ 
and $f(R)$ gravities this is not valid anymore, due to the boundary term. 

In this work  we are interested in $f(Q,\mathcal{T})$ gravity, which is another 
possible extension of STEGR 
\cite{Xu:2019sbp,Yang:2021fjy,Najera:2021afa,Arora:2020tuk,Arora:2020met,
Gadbail:2023suo,Gadbail:2023klq,Gadbail:2022qnw,Gadbail:2021fjf}, with 
five-dimensional gravitational action given by 
\begin{equation}\label{a1}
S=\int d^5x \sqrt{-g}\Big[\frac{1}{4}f(Q,\mathcal{T})+\mathcal{L}_m\Big],
\end{equation}
where $\mathcal{L}_m$ represents  the matter lagrangian to be defined in the 
next section. The variation of action (\ref{a1}) with respect to the metric 
yields the following modified Einstein equation
\begin{equation}
G_{MN}=2\Big[\mathcal{T}_{MN}-\frac{f_T}{2}(\mathcal{T}_{MN}+\theta_{MN})\Big],  
\end{equation}
where
\begin{equation}
G_{MN}= \frac{2}{\sqrt{-g}}\nabla_K(\sqrt{-g}f_QP^K\ _{MN}) -\frac{1}{2}g_{MN}f+f_Q(P_{MKL}Q_N\ ^{KL}-2Q^ L\ _{KM}P^K\ _{NL}).  
\end{equation}
In addition, $\mathcal{T}_{MN}$ is the  momentum-energy tensor given by
\begin{equation}
\mathcal{T}_{MN}=-2\frac{\delta \mathcal{L}_m}{\delta g^{MN}}+ g_{MN}\mathcal{L}_m,
\end{equation}
while the tensor $\theta_{MN}$ is defined as
\begin{equation}
\theta_{MN}=g^{AB}\frac{\delta T_{AB}}{\delta g^{MN}}.
\end{equation}

We can still vary the action (\ref{a1}) with respect to the connection to obtain
\begin{equation}
\nabla_M\nabla_N(\sqrt{-g}f_Q P_K\ ^{MN})=0.   
\end{equation}
Above, we have defined $f\equiv f(Q,T)$, $f_Q\equiv \partial f(Q,T)/\partial Q$ 
and $f_T\equiv \partial f(Q,T)/\partial T$ for simplicity.

\section{construction of the cuscuton brane}
\label{s3}

Once we have revised the basic concepts of STEGR, we are now able to 
build the braneworld scenarios. As a starting point  let us consider a single 
scalar field with noncanonical dynamics as a matter source. This scalar field 
depends only on an extra dimension and provides thickness for the brane. The 
matter Lagrangian is then given  by
\begin{equation}
\mathcal{L}_m=\mathcal{L}(X) -V(\phi),   
\end{equation}
where $X=-\partial_M\phi\partial^M\phi/2$. The momentum-energy tensor associated 
with this Lagrangian is 
\begin{equation}
\mathcal{T}_{MN}=\mathcal{L}_X\partial_M\phi\partial_N\phi+g_{MN}\mathcal{L}_m,    
\end{equation}
while its trace reads
\begin{equation}
\mathcal{T}=\mathcal{L}_X\partial_M\phi\partial^M\phi+5\mathcal{L}(X)-5V. 
\end{equation}
Thus, we write the tensor $\theta$ as   
\begin{equation}
\theta_{MN}= -\Big(\frac{1}{2}\mathcal{L}_{XX}\partial_A\phi\partial^A\phi+\frac{5}{2}\mathcal{L}_X\Big)\partial_M\phi\partial_N\phi.   
\end{equation}

As a concrete noncanonical scalar field, we will consider the standard kinetic 
term along with a cuscuton term, whose strength is controlled by 
the non-negative parameter $\alpha$. The lagrangian for this  noncanonical 
scalar field reads as \cite{Afshordi:2006ad,Afshordi:2007yx}
\begin{equation}
\mathcal{L}(X)=X+\alpha\sqrt{X}. 
\end{equation}
Hence, for $\alpha=0$, the standard dynamics is recovered.

Let us now investigate how the cuscuton term affect the braneworld scenarios. 
To obtain the brane equations, one considers a flat Randall-Sundrum-like metric, 
written as
\begin{equation}\label{metric}
ds^2= e^{2A}\eta_{\mu\nu}dx^{\mu}dx^{\nu}+dy^2,    
\end{equation}
where $\eta_{\mu\nu}$ is the Minkowski metric, $e^{2A}$ is the warp factor  and 
$y$ represent the extra dimension. Like the scalar field, $A(y)$ is  a
function only of the extra dimension. Note that the Greek indices $\mu,\nu$ 
run from 0 to 3. Besides, we should point out that the coincident gauge, i.e., 
$\widetilde{\Gamma}^P\ _{MN}=0$ is considered. Then, for this metric, the scalar 
field and gravitational equations read
\begin{eqnarray}
\Big(1+\frac{3}{4}f_{\mathcal{T}}\Big)\phi^{\prime\prime}+\Big[(4+3f_{\mathcal{T
}})A^{\prime}+\frac{3}{4} f_{\mathcal{T}}^{\prime}\Big]\phi^{\prime} 
-[4(1+f_T)A^{\prime}+f_{\mathcal{T}}^{\prime}]\alpha&=&\Big(1+\frac{5}{4}f_{
\mathcal{T}}\Big)V_{\phi},\\
\label{erer}
12f_Q A^{\prime 2}-\frac{1}{2}f&=&\Big(1+\frac{3}{2}f_{\mathcal{T}}\Big)\phi^{\prime 2}-2\alpha f_{\mathcal{T}}\phi^{\prime}-2V,\\
\label{erer2}
3(f_Q A^{\prime\prime}+f_Q^{\prime}A^{\prime })&=&-\Big(2+\frac{3f_{\mathcal{T}}}{2}\Big)\phi^{\prime 2}+2\alpha(1+f_{\mathcal{T}})\phi^{\prime}.
\end{eqnarray}
In the above equations, the prime ($^{\prime}$)  denotes derivative with respect 
to the extra dimension. To solve the these equations  we introduce the 
first-order formalism by writing the derivative of the warp factor with respect 
to the extra dimension as a function of scalar field, namely
\cite{Gremm1999,Menezes,Moreira:2021uod}
\begin{equation}\label{A}
A^{\prime}=-a W(\phi),    
\end{equation}
where $W(\phi)$ is the superpotential. Concerning the   function 
$f(Q,\mathcal{T})$ in the following  
we consider the quite general power-law     form 
\begin{equation}
 f(Q,\mathcal{T})=Q+k_1Q^n+k_2 \mathcal{T},
 \label{powerlawform}
\end{equation}
where $k_1$ controls the effect of nonmetricity and $k_2$ controls the 
effect of trace of momentum-energy tensor. In addition, we have 
$Q=12A^{\prime 2}$, and thus   Eqs.(\ref{erer} and \ref{erer2}) are rewritten 
as
\begin{eqnarray}
\label{escalar}
&&\phi^{\prime}=\frac{2}{4+3k_2}\Big\{2\alpha(1+k_2)+3a W_\phi\Big[1+nC_n(a 
W)^{2n-2}\Big]\Big\},\\
&&V(\phi)=\frac{4}{4+5k_2}\Bigg\{\frac{2+k_2}{(4+3k_2)^2} 
\Bigg[2\alpha(1+k_2)+3a W_\phi\Big[1+nC_n(a 
W)^{2n-2}\Big]\Bigg]^2 -3(aW)^2[1+C_n(aW)^{2n-2}]\Bigg\},
\end{eqnarray}
where $C_n=12^{n-1}k(2n-1)$.

The energy density is given by
\begin{eqnarray}
 &&\!\!\!\!\!\!\rho(y)=e^{2A}\Bigg\{  
\frac{2(8+11k_2)}{(4+5k_2)(4+3k_2)^2}\Bigg[2\alpha(1+k_2)+3a W_\phi\Big[1+nC_n(a 
W)^{2n-2}\Big]\Bigg]^2\nonumber\\&&\ \ \ \ \ \ -\frac{2\alpha}{4+3k_2} 
\Bigg[2\alpha(1+k_2)+3a W_\phi\Big[1+nC_n(a 
W)^{2n-2}\Big]\Bigg]-\frac{12(aW)^2}{4+5k_2}\Big[1+nC_n(a 
W)^{2n-2}\Big]\Bigg\}.
\end{eqnarray}

The braneworld system is  described by scalar field solution $\phi$, warp factor 
$A(y)$, potential $V(\phi)$, and energy density $\rho(y)$, which are completely 
defined by superpotential. The next step is to choose a specific form for 
$W(\phi)$. A simple example is the sine-Gordon with $n=1$.

\begin{figure}[ht!]
\begin{center}
\begin{tabular}{ccc}
\includegraphics[height=4.5cm]{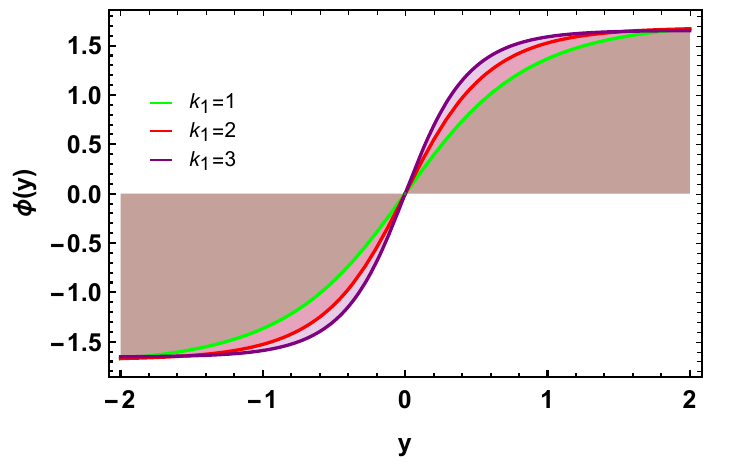} 
\includegraphics[height=4.5cm]{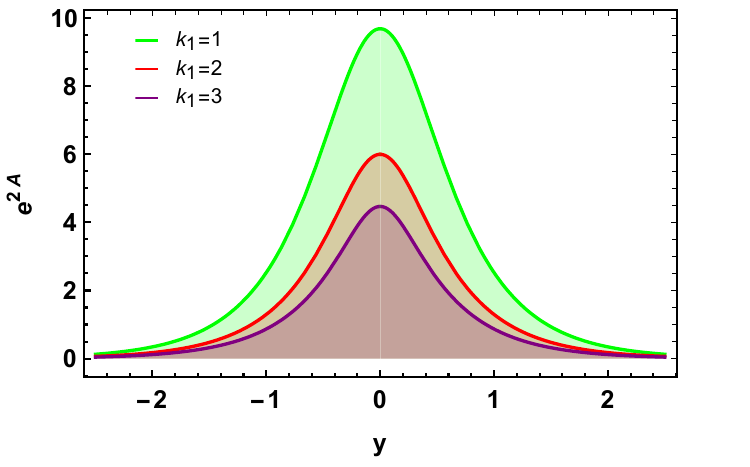}\\
(a)\hspace{6.7cm}(b)\\
\includegraphics[height=4.5cm]{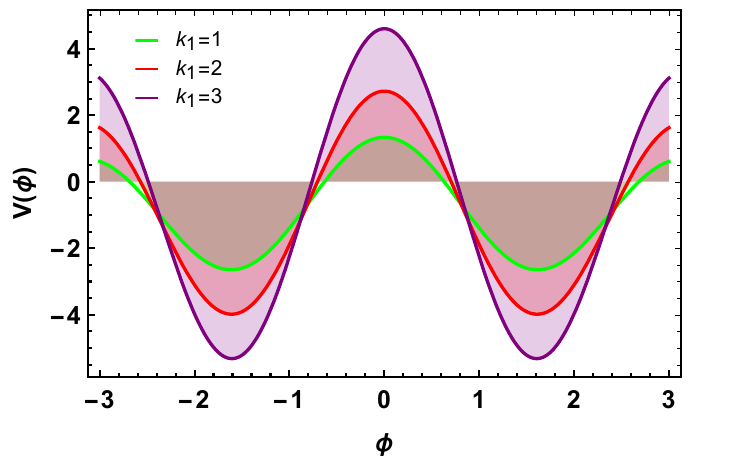}
\includegraphics[height=4.5cm]{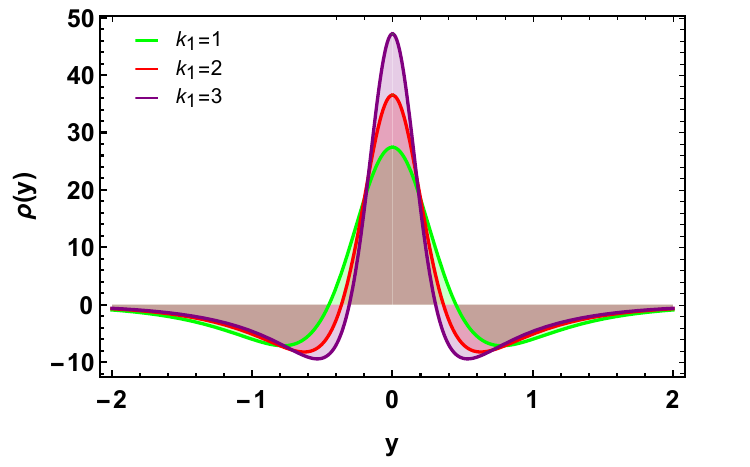}\\
(c)\hspace{6.7cm}(d) 
\end{tabular}
\end{center}
\vspace{-0.5cm}
\caption{ The case of the sine-Gordon superpotential (\ref{SinGordon}), with 
$a=\beta=k_2=1$, $\alpha=0,25$  and varying $k_2$. (a) Scalar field. (b) 
Warp factor. (c) Potential. (d) Energy density.
\label{fig1}}
\end{figure}

\begin{figure}[ht!]
\begin{center}
\begin{tabular}{ccc}
\includegraphics[height=4.5cm]{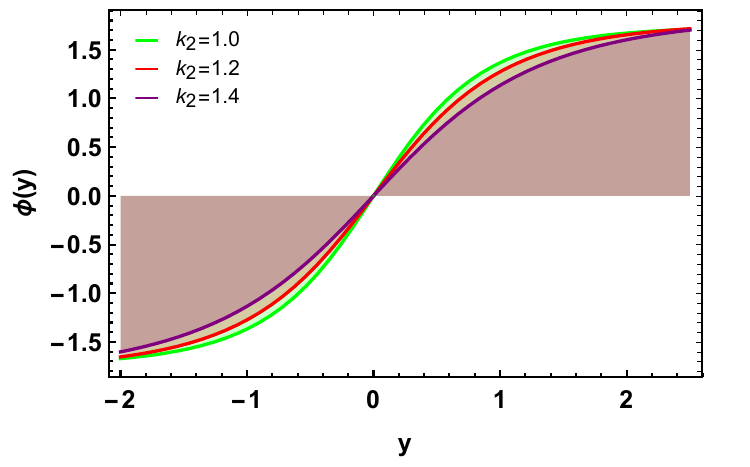} 
\includegraphics[height=4.5cm]{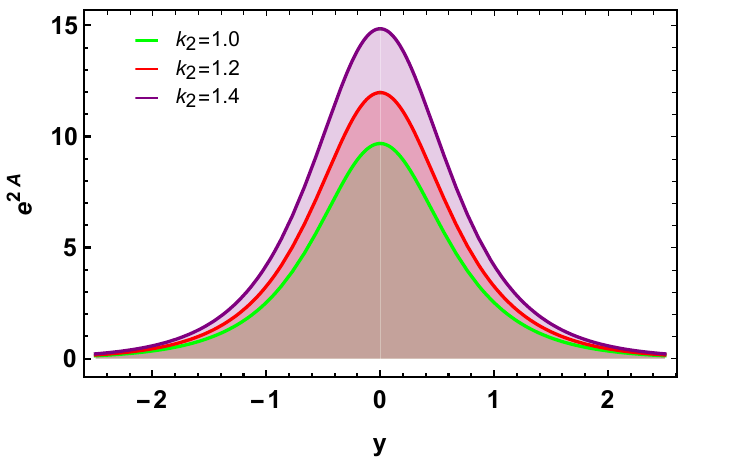}\\
(a)\hspace{6.7cm}(b)\\
\includegraphics[height=4.5cm]{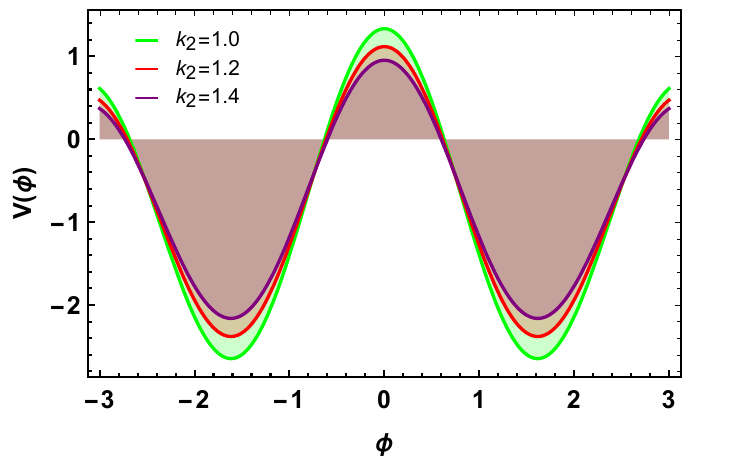}
\includegraphics[height=4.5cm]{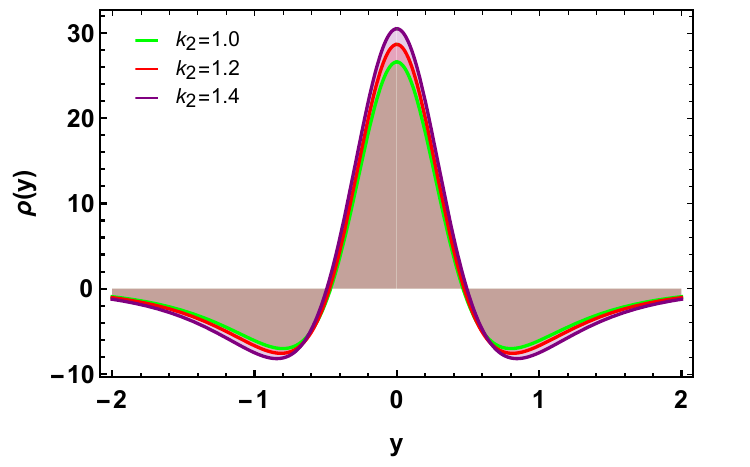}\\
(c)\hspace{6.7cm}(d) 
\end{tabular}
\end{center}
\vspace{-0.5cm}
\caption{ The case of the sine-Gordon superpotential (\ref{SinGordon}), with 
$a=\beta=k_1=1$, $\alpha=0,25$    and varying $k_1$. (a) Scalar field. (b) 
Warp factor. (c) Potential. (d) Energy density.
\label{fig2}}
\end{figure}
\subsection{Sine-Gordon superpotential}

The specific form for the sine-Gordon superpotential is 
\begin{equation}
W(\phi)=\beta^2\sin{\Big(\frac{\phi}{\beta}\Big)}.   
\label{SinGordon}
\end{equation}
For this superpotential and considering that $n=1$ in (\ref{powerlawform}), the 
scalar field takes the form
\begin{eqnarray}
\phi(y)&=&2 \beta  \tan ^{-1}\Bigg\{\frac{3 a\beta (k_1+1)+2 \alpha  (k_2+1) 
}{\sqrt{9 a\beta^2 (k_1+1)^2-4 \alpha^2 (k_2+1)^2}} 
\times \tanh \Bigg[\frac{y \sqrt{9 a\beta^2 (k_1+1)^2-4 \alpha^2 (k_2+1)^2}}{ 
\beta  (3 k_2+4)}\Bigg]
\Bigg\},
\end{eqnarray}
which brings us to the solution for the warp factor
\begin{eqnarray}
A(y)&=&\frac{ \beta ^2 (3 k_2+4)}{6(k_1+1)} \log \Bigg\{2 \alpha  (k_2+1)+3 
a\beta  (k_1+1)
\nonumber
\\&\times&\cos \Bigg[2 \beta  \tan ^{-1}\Bigg(\frac{3 a\beta 
(k_1+1)+2 \alpha  (k_2+1) }{\sqrt{9 a\beta^2 (k_1+1)^2-4 \alpha^2 
(k_2+1)^2}} 
 \times  \tanh \Bigg(\frac{y \sqrt{9 a\beta^2 (k_1+1)^2-4 \alpha^2 (k_2+1)^2}}{ 
\beta  (3 k_2+4)}\Bigg)
\Bigg)\Bigg]\Bigg\}.
\end{eqnarray}
Furthermore, the potential takes the form
\begin{eqnarray}
V(\phi)=\frac{1}{4+5k_2}\Bigg\{\frac{4(2+k_2)}{(4+3k_2)^2}\Big[
2\alpha(1+k_2)+3a \beta(1+k_1)\cos\Big(\frac{\phi}{\beta}\Big)\Big]^2 
-12(1+k_1)a^2\beta^2\sin\Big(\frac{\phi}{\beta}\Big)^2\Bigg\}.
\end{eqnarray}
The solutions for the scalar field obey the form of a kink-like structure, 
and as expected  the potential has an oscillating form. This brings us to a 
localized energy density for the brane. 

Furthermore, it is interesting to analyze the effect of the parameters 
$k_{1,2}$ on the brane solutions. Note that by increasing the value of the 
nonmetricity parameter $k_1$, the kink-like field intensifies, the warp factor 
decreases its amplitude, the potential increases the amplitude of its 
oscillations, and the density becomes more localized (see Fig. \ref{fig1}). On 
the other hand, when we increase the value of the momentum-energy parameter 
$k_2$, the amplitude of the warp factor is intensified, and the potential has 
the amplitudes of its oscillations reduced (see Fig. \ref{fig2}). Despite this 
behavior, contrary to the variation of $k_1$  when we increase the value of 
$k_2$ the energy density becomes more localized.

\begin{figure}[ht!]
\begin{center}
\begin{tabular}{ccc}
\includegraphics[height=4.5cm]{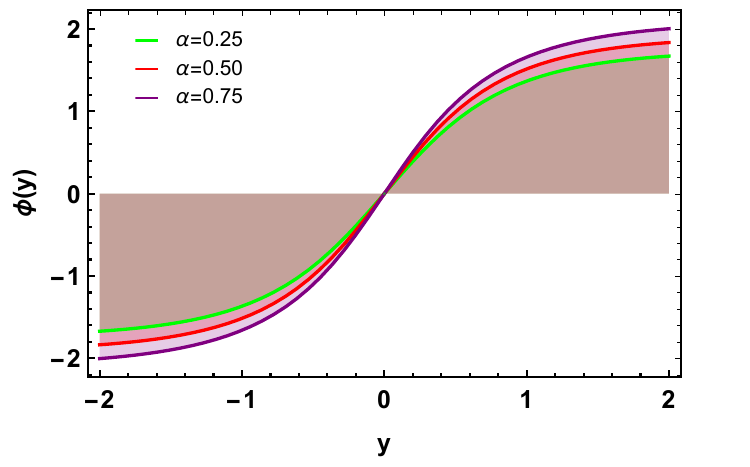} 
\includegraphics[height=4.5cm]{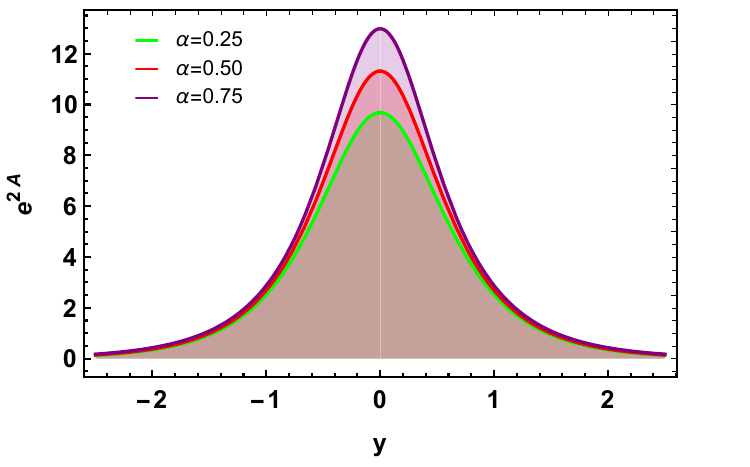}\\
(a)\hspace{6.7cm}(b)\\
\includegraphics[height=4.5cm]{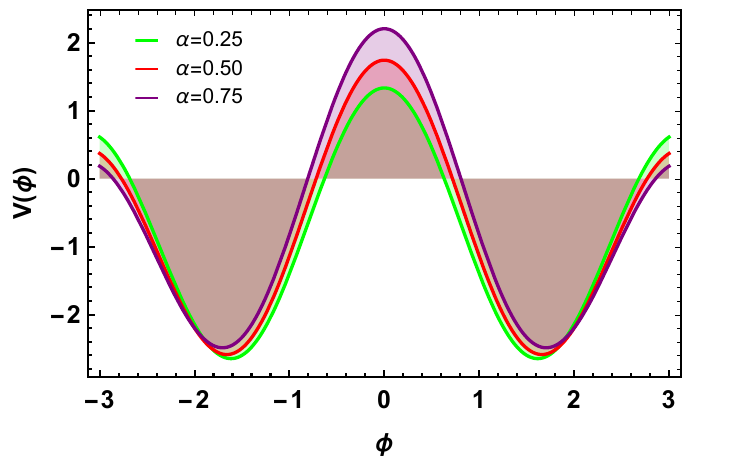}
\includegraphics[height=4.5cm]{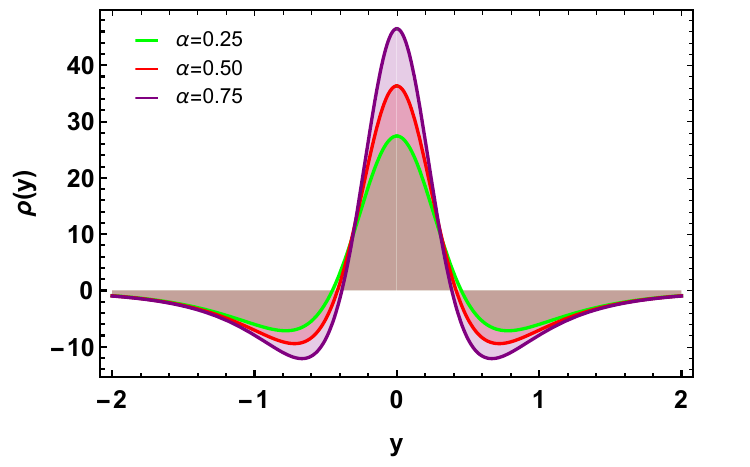}\\
(c)\hspace{6.7cm}(d) 
\end{tabular}
\end{center}
\vspace{-0.5cm}
\caption{ The case of the sine-Gordon superpotential (\ref{SinGordon}), with 
$a=\beta=k_1=k_2=1$, and   varying  $\alpha$. (a) Scalar field. (b) Warp 
factor. (c) Potential. (d) Energy density.
\label{fig14}}
\end{figure}

The cuscuton  parameter   affects the brane solutions, too.  As the behavior is 
similar for any chosen superpotential, in Fig.  \ref{fig14} we depict the 
graphical behavior of the scalar field, warp factor, potential and energy 
density. We can notice that the kink-like solution intensifies when we increase 
the value of $\alpha$, increasing the amplitude of the warp factor. 
Additionally, the potential increases the amplitude of its oscillations and the 
energy density becomes more localized.

\subsection{Polynomial superpotential}

Let us now investigate another specific example, namely the polynomial 
superpotential, which has the form
\begin{equation}
 W(\phi)=\beta\phi-\frac{1}{3\beta}\phi^3.  
 \label{Polynomialsuper}
\end{equation}
By considering   $n=1$ in (\ref{powerlawform}), the scalar field becomes
\begin{eqnarray}
\phi(y)=\sqrt{\frac{3 a \beta^2  (k_1+1)+2 \alpha\beta  (k_2+1)}{3a(k_1+1)}} 
\times\tanh \Bigg[\frac{2 \sqrt{3 a (k_1+1) [ 3 a \beta  (k_1+1)+2 \alpha  
(k_2+1)]}y}{\sqrt{\beta } (3 k_2+4)}\Bigg].
\end{eqnarray}
In turn, the warp factor becomes
\begin{eqnarray}
&&
\!\!\!\!\!\!\!\!\!\!\!\!\!\!\!\!
A(y)=\frac{\beta  (3 k_2+4)}{108 a (k_1+1)^2}\Bigg\{4 \Big[\alpha  (k_2+1)-3 
a \beta  (k_1+1)\Big] \times \ln \Bigg[\cosh \Bigg(\frac{2 \sqrt{3 a 
(k_1+1) [ 3 a \beta  (k_1+1)+2 \alpha  (k_2+1)]}y}{\sqrt{\beta } (3 
k_2+4)}\Bigg)\Bigg]
\nonumber\\
&&
\ \ \ \ \ \ \ \ \ \ \ \ \ \ \ \ \ \ \ \ \ \ \ 
-\Big[3 a \beta  (k_1+1)+2 \alpha  
(k_2+1)\Big]  \times \tanh ^2\Bigg(\frac{2 \sqrt{3 a (k_1+1) [ 3 a 
\beta  (k_1+1)+2 \alpha  (k_2+1)]}y}{\sqrt{\beta } (3 k_2+4)}\Bigg)\Bigg\}.    
\end{eqnarray}
Thus,  the potential becomes
\begin{eqnarray}
V(\phi)=\frac{1}{4+5k_2}\Bigg\{ 
\frac{4(2+k_2)}{(4+3k_2)^2}\Big[2\alpha(1+k_2)+3a 
(1+k_1)\Big(\beta-\frac{1}{\beta}\phi^2\Big)\Big]^2
 -12a^2(1+k_1)\Big(\beta\phi-\frac{1}{3\beta}\phi^3\Big)^2\Bigg\}.
\end{eqnarray}

\begin{figure}[ht!]
\begin{center}
\begin{tabular}{ccc}
\includegraphics[height=4.5cm]{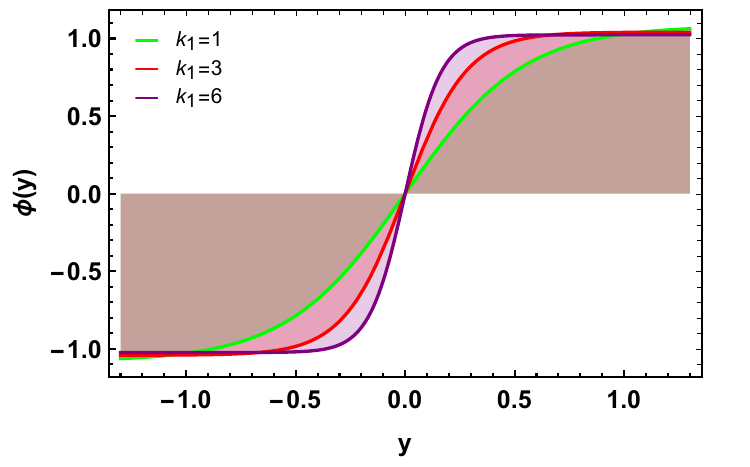} 
\includegraphics[height=4.5cm]{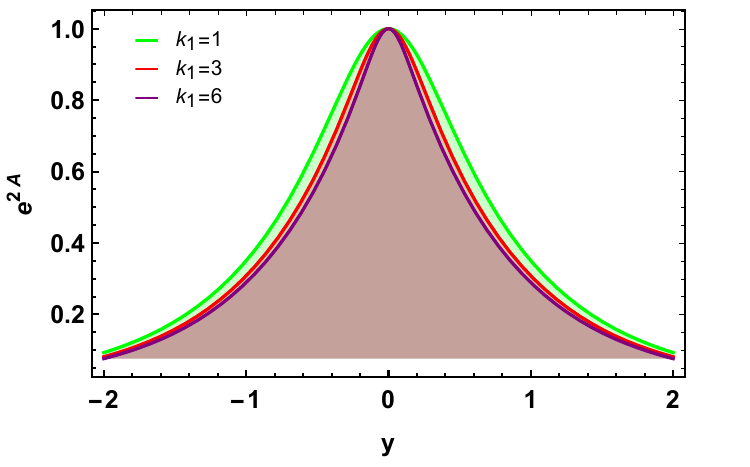}\\
(a)\hspace{6.7cm}(b)\\
\includegraphics[height=4.5cm]{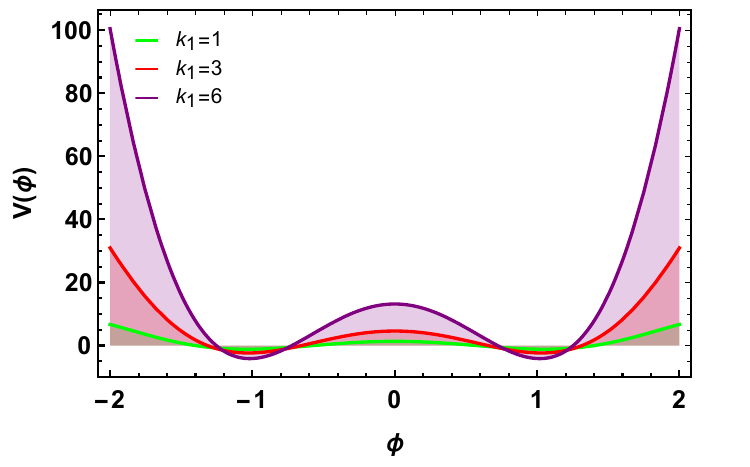}
\includegraphics[height=4.5cm]{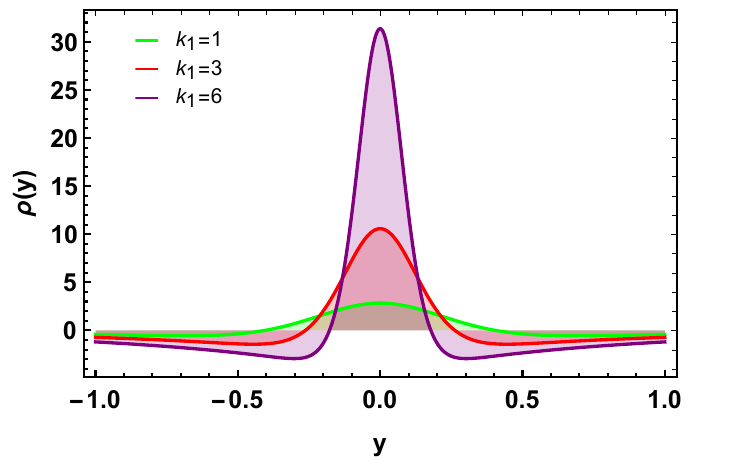}\\
(c)\hspace{6.7cm}(d) 
\end{tabular}
\end{center}
\vspace{-0.5cm}
\caption{ The case of the polynomial superpotential (\ref{Polynomialsuper}), 
with $a=\beta=k_2=1$, $\alpha=0,25$, and varying $k_1$. (a) Scalar field. (b) 
Warp factor. (c) Potential. (d) Energy density.
\label{fig3}}
\end{figure}

\begin{figure}[ht!]
\begin{center}
\begin{tabular}{ccc}
\includegraphics[height=4.5cm]{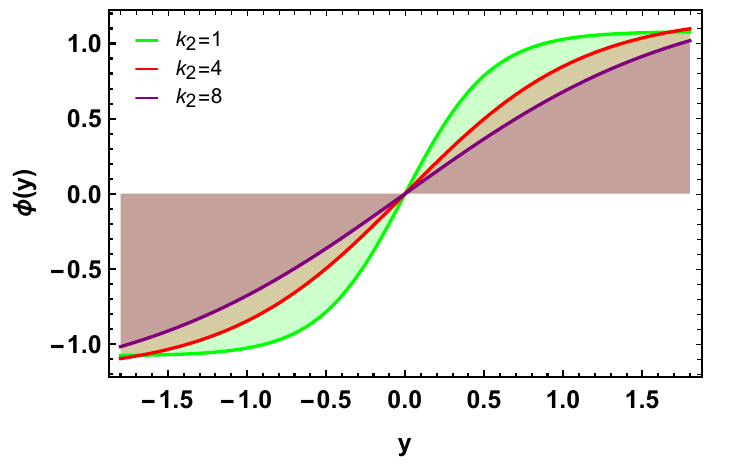} 
\includegraphics[height=4.5cm]{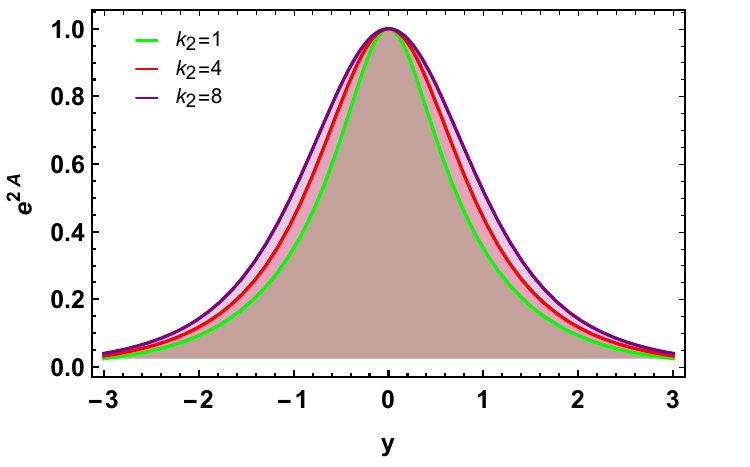}\\
(a)\hspace{6.7cm}(b)\\
\includegraphics[height=4.5cm]{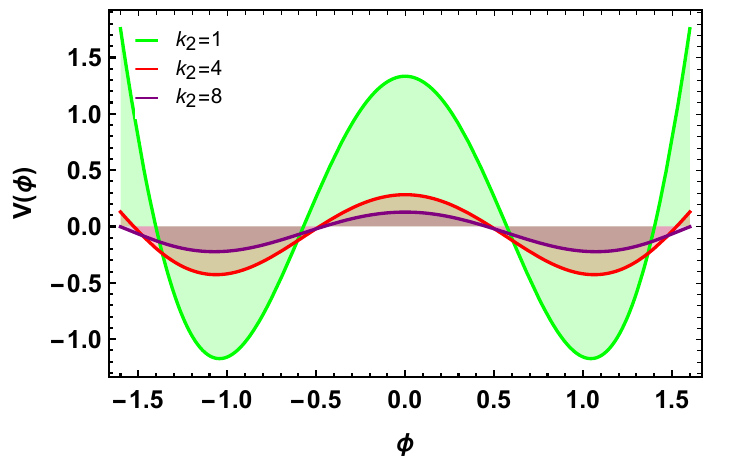}
\includegraphics[height=4.5cm]{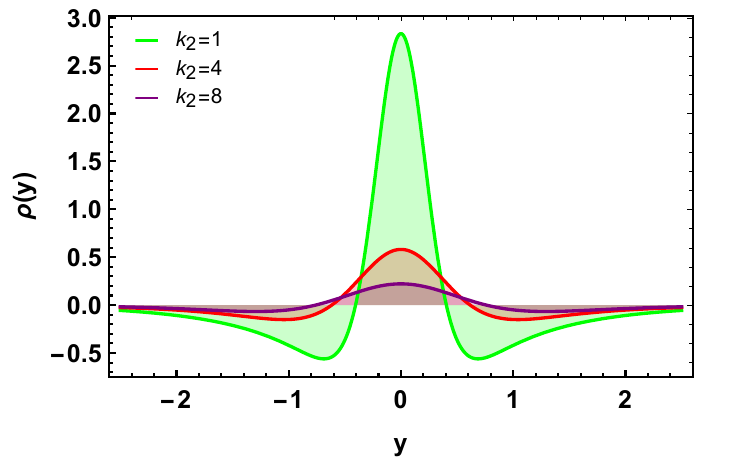}\\
(c)\hspace{6.7cm}(d) 
\end{tabular}
\end{center}
\vspace{-0.5cm}
\caption{  The case of the polynomial superpotential (\ref{Polynomialsuper}),  
with $a=\beta=k_1=1$, $\alpha=0,25$ and varying $k_2$. (a) Scalar field. (b) 
Warp factor. (c) Potential. (d) Energy density.
\label{fig4}}
\end{figure}

It is worth mentioning the effect of the parameters $k_{1,2}$ on the brane 
solutions. Note that as we increase the value of the $k_1$ parameter, the 
scalar field, which is shaped like a kink-like structure, tends to intensify. 
The warp factor maintains its amplitude, however it becomes more localized as 
$k_1$ increases. This leads us to an energy density that tends to become more 
localized as $k_1$ increases (see Fig. \ref{fig3}).
On the other hand, if we decrease the value of parameter $k_2$, the kink-like 
solution intensifies, the warp factor becomes more localized and the energy 
density presents a more localized profile, with a very accentuated peak at the 
origin (see Fig. \ref{fig4}).

 \begin{figure}[ht!]
\begin{center}
\begin{tabular}{ccc}
\includegraphics[height=4.5cm]{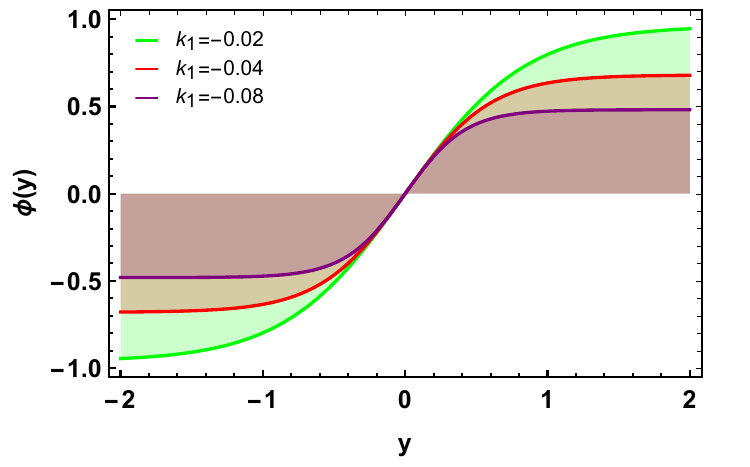} 
\includegraphics[height=4.5cm]{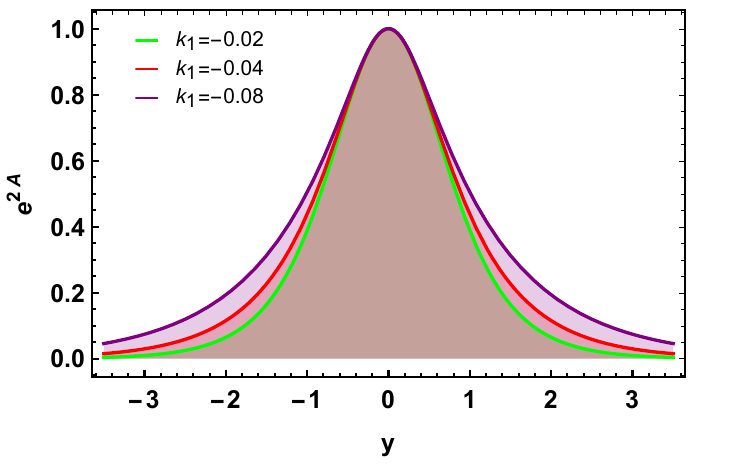}\\
(a)\hspace{6.7cm}(b)\\
\includegraphics[height=4.5cm]{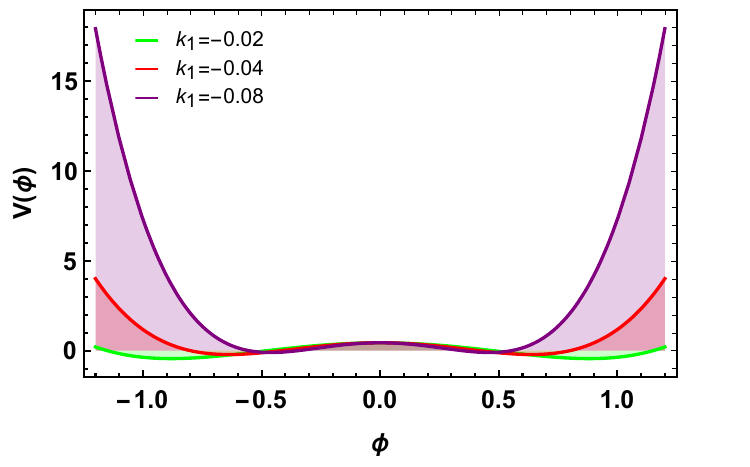}
\includegraphics[height=4.5cm]{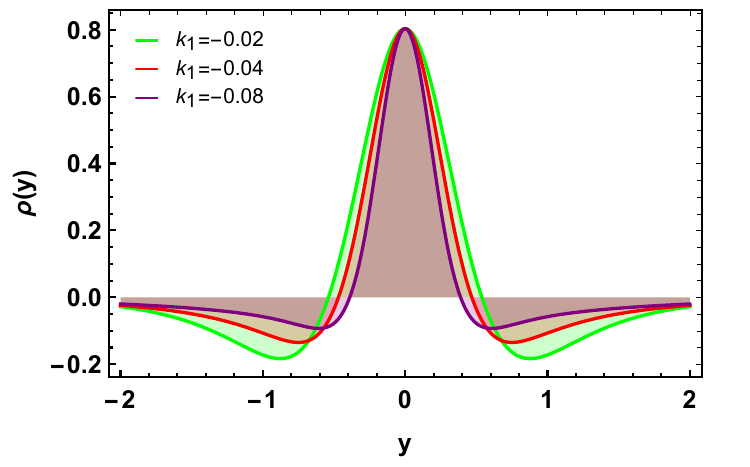}\\
(c)\hspace{6.7cm}(d) 
\end{tabular}
\end{center}
\vspace{-0.5cm}
\caption{  The case of the linear superpotential (\ref{Linearsuper}),  
 with $a=\beta=k_2=1$, $\alpha=0,25$, and varying $k_1$. (a) 
Scalar field. (b) Warp factor. (c) Potential. (d) Energy density.
\label{fig5}}
\end{figure}
\subsection{Linear superpotential}

We now consider the case of the
linear superpotential 
\begin{eqnarray}
W(\phi)=\beta\phi.
 \label{Linearsuper}
\end{eqnarray}
Let us consider $n=2$ in    (\ref{powerlawform}).
For these choices,   the scalar field is written as
\begin{eqnarray}
\phi(y)=\frac{\sqrt{3  a\beta +2 \alpha  (k_2+1)}}{6a\beta \sqrt{6k_1 a\beta}} 
\tan \Bigg[\frac{12a\beta \sqrt{6k_1 a\beta[3 a\beta +2 \alpha  (k_2+1)]}y}{3 
k_2+4}\Bigg], 
\end{eqnarray}
and the warp factor reads as
\begin{eqnarray}
A(y)=\frac{3 k_2+4}{432 a^2\beta^2  k_1} \ln \Bigg[\cos \Bigg(\frac{12a\beta 
\sqrt{6k_1 a\beta(3 a\beta +2 \alpha  (k_2+1))}y}{3 k_2+4}\Bigg)\Bigg].
\end{eqnarray}
Additionally, the potential becomes
\begin{eqnarray}
V(\phi)=\frac{1}{4+5k_2}\Bigg\{ 
\frac{4(2+k_2)}{(4+3k_2)^2}\Big[2\alpha(1+k_2)+3a \beta(1+72 
k_1a^2\beta^2\phi^2)\Big]
^2 -12a^2\beta^2\phi^2(1+36k_1a^2\beta^2\phi^2)\Bigg\}.
\end{eqnarray}

\begin{figure}[ht!]
\begin{center}
\begin{tabular}{ccc}
\includegraphics[height=4.5cm]{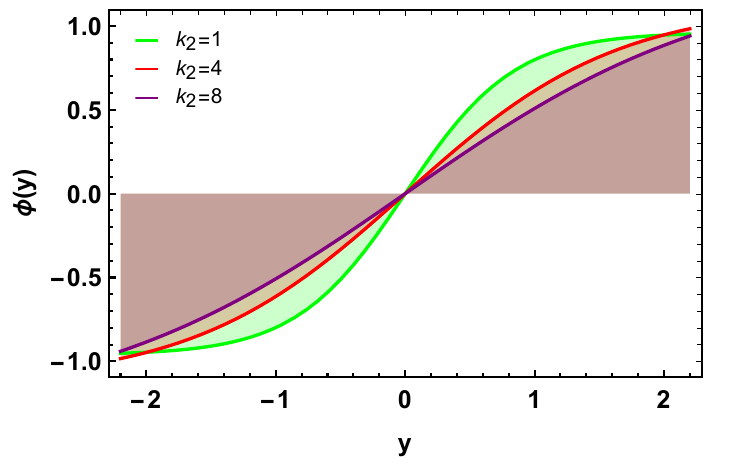} 
\includegraphics[height=4.5cm]{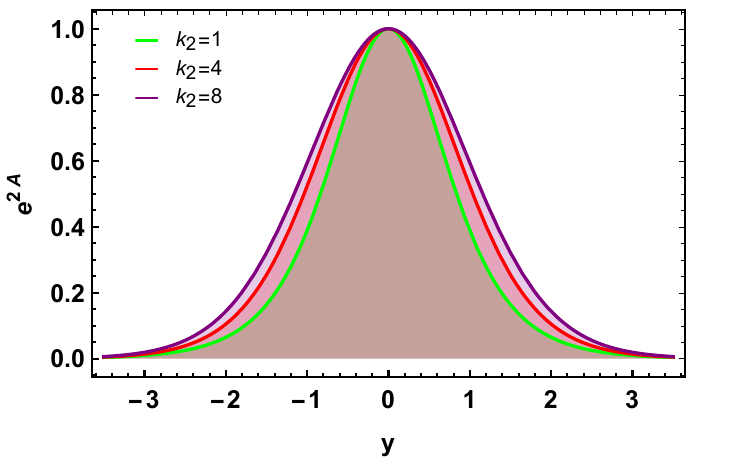}\\
(a)\hspace{6.7cm}(b)\\
\includegraphics[height=4.5cm]{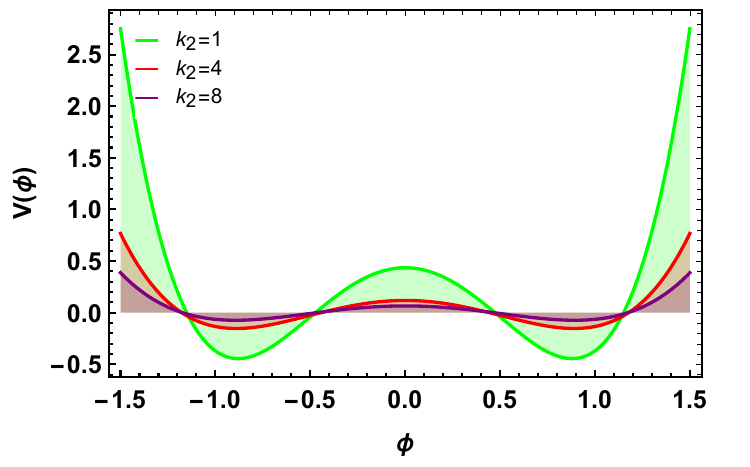}
\includegraphics[height=4.5cm]{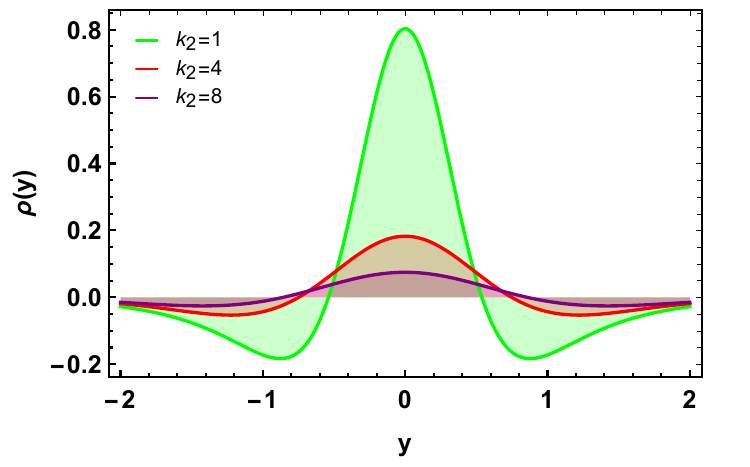}\\
(c)\hspace{6.7cm}(d) 
\end{tabular}
\end{center}
\vspace{-0.5cm}
\caption{ The case of the linear superpotential (\ref{Linearsuper}),   with 
$a=\beta=1$, $k_1=-0.02$, $\alpha=0,25$, and varying $k_2$. (a) Scalar field. 
(b) Warp factor. (c) Potential. (d) Energy density.
\label{fig6}}
\end{figure}

As we observe, the energy density is well located on the brane. The scalar 
field solution has the behavior of a kink-like structure. Furthermore, an 
analysis of the effect of the parameters $k_{1,2}$ on the brane solutions 
reveals interesting results. When the value of the parameter $k_1$ is 
decreased, the kink-like solution tends to become more compact, and this 
behavior is reflected in the potential and in the warp factor, which becomes 
less localized. On the other hand, the energy density  becomes more 
localized (see Fig. \ref{fig5}).
Finally, when we decrease the value of the parameter $k_2$, the kink-like 
solution becomes more intense, the warp factor becomes more localized, and the 
energy density increases its amplitude becoming more localized 
(see Fig. \ref{fig6}).

\section{Probabilistic measures for fermion location}
\label{s4}

In the previous section we examined in detail how the parameters that control 
deviations from standard STEGR gravity ($k_{1,2}$ and $n$), as well as the  
cuscuton dynamics ($\alpha$), affect the structure of the brane. To deepen our 
understanding, this section covers the localization of spin-$1/2$ fermions. 
This is a very interesting subject, since for fermions to be successfully 
located on the brane it is essential to establish a coupling between the 
fermionic field and the background scalar field ($\phi$). This type of 
interaction is known as Yukawa-type coupling, represented by 
$\overline{\Psi}\phi\Psi$. This coupling mechanism has been widely used in 
several studies due to its simplicity and effectiveness in accurately describing 
the dynamics of fermions \cite{Liu2008, Liu2008b, Liu2009, Liu2009a, Liu2009b, 
Dantas2013, Moreira20211}. Thus, when adopting a minimal Yukawa-type coupling, 
the action represented by the spin-$1/2$ Dirac field in a five-dimensional 
spacetime is expressed as
\begin{eqnarray}\label{1}
\mathcal{S}_{1/2}=\int \sqrt{-g} \Big(\overline{\Psi}i\Gamma^M D_M\Psi -\xi 
\phi\overline{\Psi}\Psi\Big)d^5x,
\end{eqnarray}
where $\xi$ represents a dimensionless coupling constant. Additionally, we have 
introduced  the covariant derivative $D_M$,  given by $D_M = \partial_M + 
\Omega_M$, qith $\Omega_M$ being the torsion-free spin connection. This 
connection is defined in   terms of the Levi-Civita connection as
\begin{eqnarray}\label{3}
\Omega_M=\frac{1}{4}\Big(\Gamma_M\ ^{{\overline{N}}{\overline{Q}}}\Big)\ 
\Gamma_{\overline{N}}\Gamma_{\overline{Q}},
\end{eqnarray}
where the curved Dirac matrices $\Gamma^M$ are obtained from the flat Dirac 
matrices  $\Gamma^{\overline{M}}$ and the  vielbeins fields $E_{\overline{M}}\ 
^M$ (the uppercase Latin indices with slash, $\overline{M}=0,...,D-1$, 
represent the tangent space coordinates), in the form:
$\Gamma^M=E_{\overline{M}}\ ^M \Gamma^{\overline{M}}$, while these matrices 
satisfy the Clifford algebra $\{\Gamma^M,\Gamma^N\}=2g^{MN}$. 
As usual, the  vielbeins  are the basis on the tangent space, and are related 
to the metric through the expression 
\begin{eqnarray}
g_{MN} = \eta_{\overline{M}\overline{N}} E^{\overline{M}}_M E^{\overline{N}}_N.
\end{eqnarray}

To simplify our study, we apply the transformation $dz = e^{-A(y)} dy$. Thus, 
the Dirac equation (\ref{1}) can be expressed as
\begin{eqnarray}\label{7}
\Big[\gamma^{\mu}\partial_\mu+\gamma^4(\partial_z+2\dot{A})-\xi 
e^A\phi\Big]\psi=0,
\end{eqnarray}
where (\ $\dot{ }$\ ) denotes a derivative with respect to $z$.

\begin{figure}[ht!]
\begin{center}
\begin{tabular}{ccc}
\hspace{-1.3cm}
\includegraphics[height=4.cm]{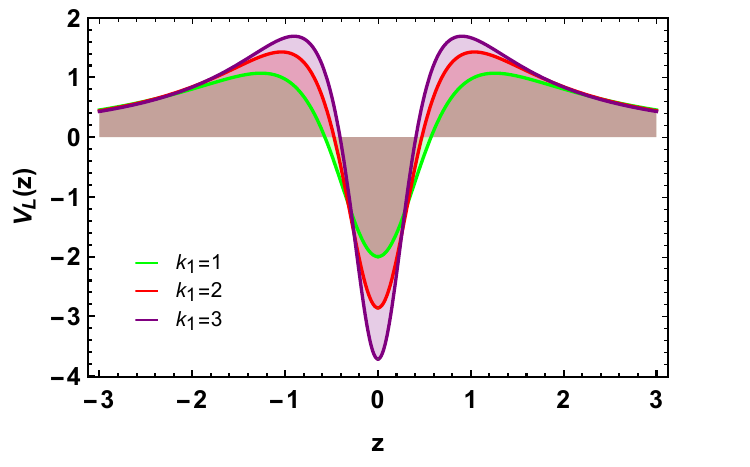} \hspace{1cm}
\includegraphics[height=4.cm]{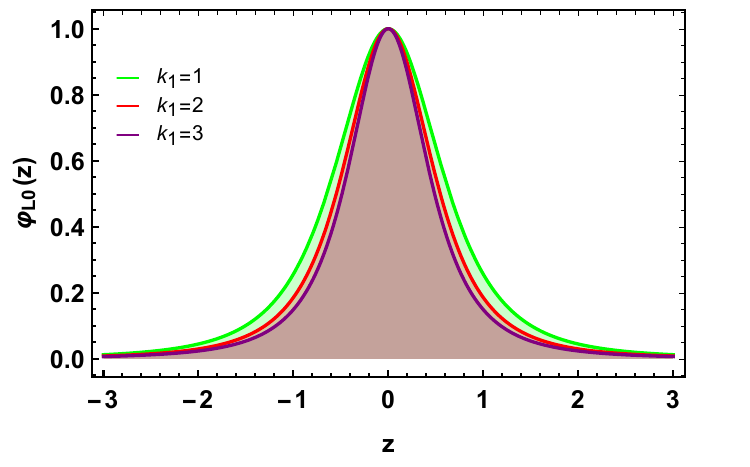}\\
(a)\hspace{6.7cm}(b)\\
\includegraphics[height=4.2cm]{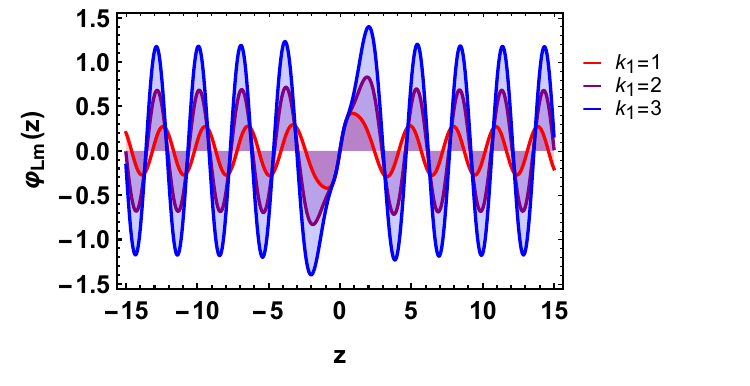}\hspace{-0.5cm}
\includegraphics[height=4.2cm]{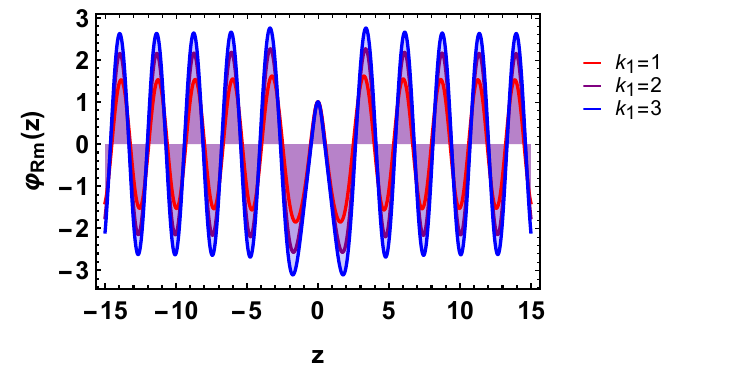}\\
(c)\hspace{6.7cm}(d) 
\end{tabular}
\end{center}
\vspace{-0.5cm}
\caption{ The case of  the sine-Gordon superpotential  (\ref{SinGordon}) with 
$a=\beta=k_2=1$, $\alpha=0,25$, and varying $k_1$. 
(a) Effective potential. (b) Massless mode. (c) Massive odd mode ($m^2=4.656$). 
(d) Massive even mode ($m^2=5.926$).
\label{fig7}}
\end{figure}

\begin{figure}[ht!]
\begin{center}
\begin{tabular}{ccc}
\hspace{-1.3cm}
\includegraphics[height=4.cm]{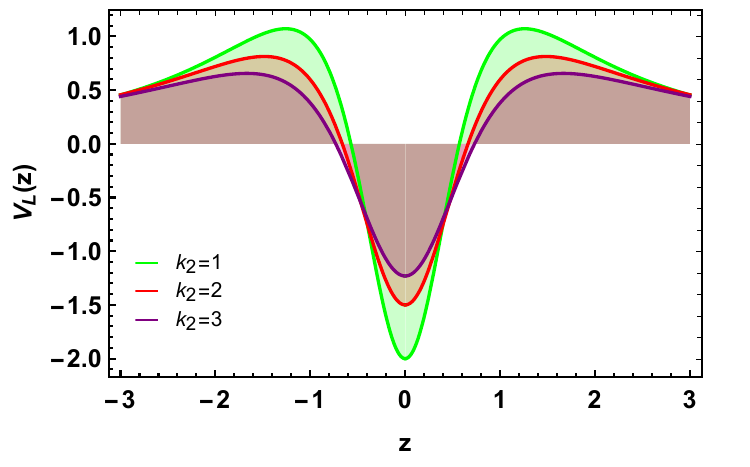} \hspace{1cm}
\includegraphics[height=4.cm]{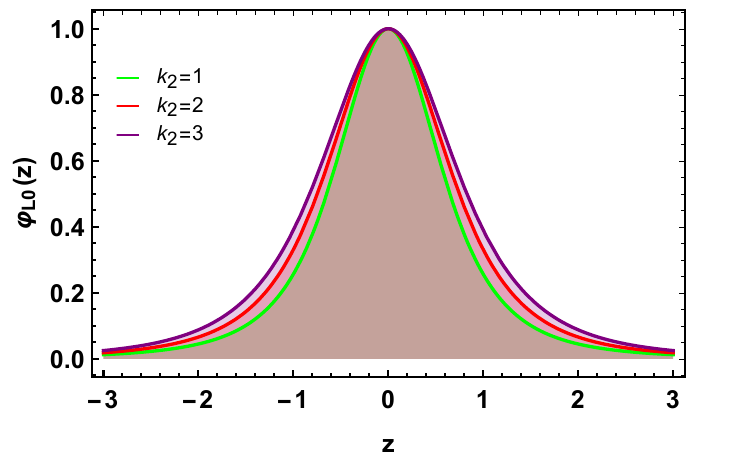}\\
(a)\hspace{6.7cm}(b)\\
\includegraphics[height=4.2cm]{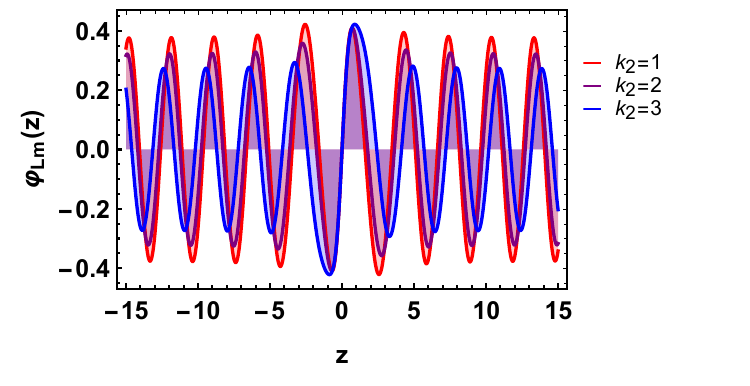}\hspace{-0.5cm}
\includegraphics[height=4.2cm]{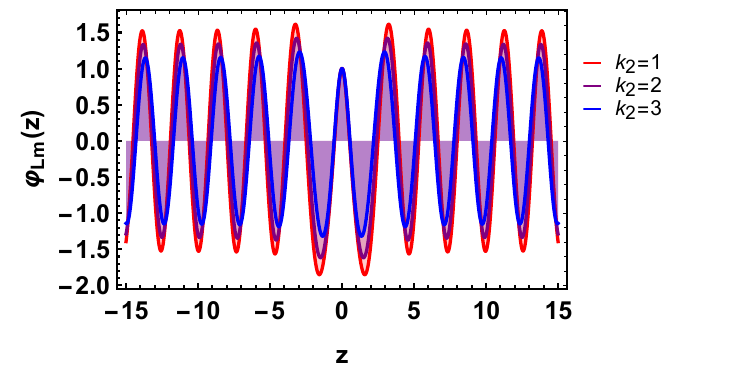}\\
(c)\hspace{6.7cm}(d) 
\end{tabular}
\end{center}
\vspace{-0.5cm}
\caption{  The case of  the sine-Gordon superpotential  (\ref{SinGordon}) with  
$a=\beta=k_2=1$, $\alpha=0,25$, and varying $k_2$. . 
(a) Effective potential. (b) Massless mode. (c) Massive odd mode ($m^2=4.656$). 
(d) Massive even mode ($m^2=5.926$).
\label{fig8}}
\end{figure}
We proceed by applying the Kaluza-Klein decomposition of the spinor
\begin{eqnarray}
\psi=\sum_n[\psi_{L,n}(x)\varphi_{L,n}(z)+\psi_{R,n}(x)\varphi_{R,n}(z)],
\end{eqnarray}
to obtain the coupled equations
\begin{eqnarray}\label{9}
\Big[\partial_z+\xi e^A \phi\Big]\varphi_{L}(z)&=&m \varphi_{R}(z),\nonumber\\
\Big[\partial_z-\xi e^A\phi\Big]\varphi_{R}(z)&=&-m \varphi_{L}(z),
\end{eqnarray}
where $\gamma^4\psi_{R,L}=\pm\psi_{R,L}$ represent the left-handed and 
right-handed components  of the Dirac field, and we also have the relation 
$\gamma^\mu\partial_\mu \psi_ {R,L}=m\psi_{L,R}$.
Equation (\ref{9}) can be decoupled, resulting to equations similar to 
Schrödinger one, namely
\begin{eqnarray}\label{10}
\Big[-\partial^2_z + V_L(z)\Big]\varphi_L(z) &=& m^2 \varphi_L(z),\nonumber\\
\Big[-\partial^2_z + V_R(z)\Big]\varphi_R(z) &=& m^2 \varphi_R(z),
\end{eqnarray}
where $V_{R,L}(z) = U^2 \pm \partial_z U$ represents  the effective potential, 
and $U = \xi e^A \phi$. It is worth mentioning that this equation takes the 
form of supersymmetric quantum mechanics (SUSY-type), guaranteeing the absence 
of Kaluza-Klein (KK) tachyon states. Furthermore, the supersymmetric structure 
allows for the existence of a well-localized massless mode of the form
\begin{eqnarray}
\varphi_{R0,L0}(z) \propto e^{\pm \int U dz}.
\end{eqnarray}
Finally, by applying boundary conditions  we can obtain the massive modes
in the form \cite{Liu2009,Liu2009a,Moreira20211,Moreira:2021wkj} 
\begin{eqnarray}\label{ade}
\varphi_{\text{even}}(0)&=&c, \quad \partial_z\varphi_{\text{even}}(0)=0, 
\nonumber\\
\varphi_{\text{odd}}(0)&=&0, \quad \partial_z\varphi_{\text{odd}}(0)=c. 
\end{eqnarray}
These boundary conditions  are chosen due to the even nature of the effective 
potentials $V_{R,L}(z)$. Moreover, the conditions (\ref{ade}) guarantee that 
solutions $\varphi_{R,L}(z)$ exhibit a behavior corresponding to even 
$\varphi_{\text{even}}$ or odd  $\varphi_{\text{odd}}$ wave functions.

In the following we examine all superpotential choices of the previous section, 
separately.

\subsection{Sine-Gordon superpotential}

\begin{figure}[ht!]
\begin{center}
\begin{tabular}{ccc}
\hspace{-1.3cm}
\includegraphics[height=4.cm]{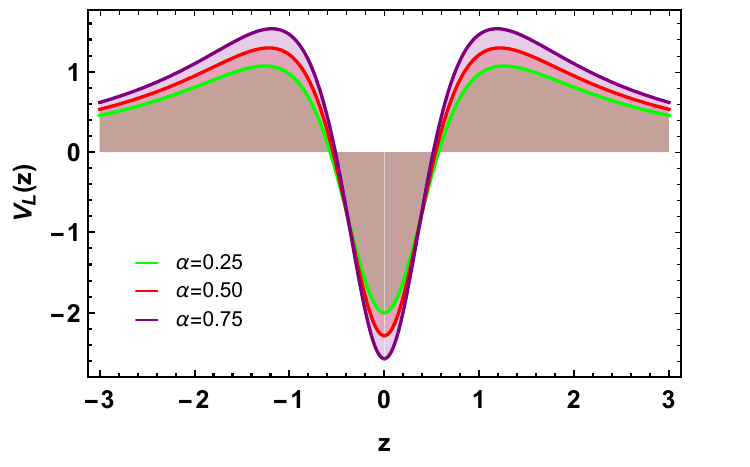} \hspace{1cm}
\includegraphics[height=4.cm]{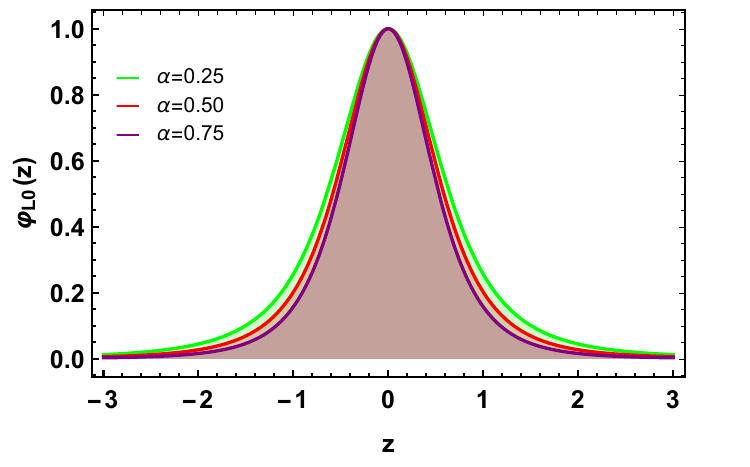}\\
(a)\hspace{6.7cm}(b)\\
\includegraphics[height=4.2cm]{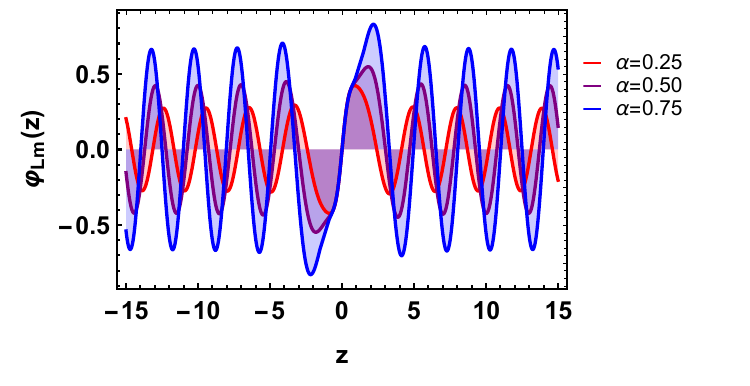}\hspace{-0.5cm}
\includegraphics[height=4.2cm]{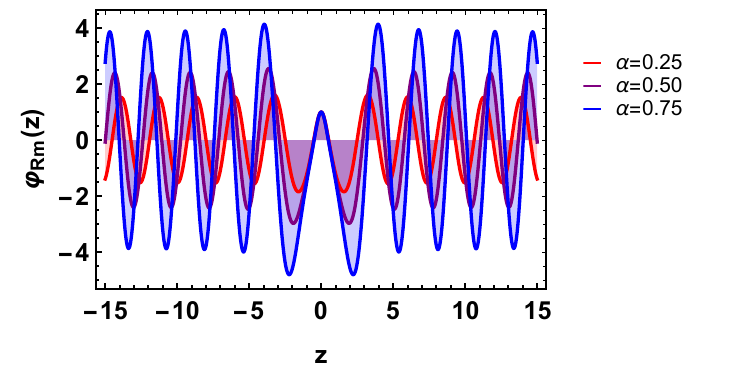}\\
(c)\hspace{6.7cm}(d) 
\end{tabular}
\end{center}
\vspace{-0.5cm}
\caption{ The case of  the sine-Gordon superpotential with $a=\beta=k_1=k_2=1$, 
and varying $\alpha$. 
(a) Effective potential. (b) Massless mode. (c) Massive odd mode ($m^2=4.656$). 
(d) Massive even mode ($m^2=5.926$).
\label{fig13}}
\end{figure}

For our first choice of the Sine-Gordon superpotential  (\ref{SinGordon}), we 
plot the graphical behaviors in 
Figs. \ref{fig7}, \ref{fig8} and \ref{fig13}. When we increase the nonmetricity 
parameter we notice that the effective potential becomes more confining, 
accentuating its well and potential barriers (see Fig. \ref{fig7} a). The 
massless 
fermionic mode feels the changes   of the effective potential, becoming 
more localized with the increase in the nonmetricity parameter $k_1$ 
(see Fig. \ref{fig7} b). Furthermore, the massive modes also feel the effect 
of 
the nonmetricity term, and thus  when we increase the    $k_1$ value we 
increase the amplitudes of the oscillations of the massive fermionic modes 
(see Fig. \ref{fig7} c and d).

On the other hand, when we increase the momentum-energy  parameter $k_2$, we 
notice that the effective potential becomes less confining (see Fig. \ref{fig8} 
a), which leads us to less localized massless fermionic modes (see 
Fig. \ref{fig8} b). This behavior is felt by massive fermionic modes that tend 
to reduce the amplitudes of their oscillations (see Fig. \ref{fig8} c and d).

The cuscuton parameter also interferes with the locations of the fermions in the 
brane.  As the behavior is similar for any chosen superpotential, we plot the 
graphical behavior of the effective potential, massless mode and massive modes 
in figure \ref{fig13}. 
When we increase the $\alpha$ parameter, we notice that the effective potential 
becomes more confining (Fig. \ref{fig13} a), making the massless fermionic mode 
more localized (Fig. \ref{fig13} b). Furthermore, massive modes increase their 
oscillation amplitudes with increasing $\alpha$ (Fig. \ref{fig13} c and d).

\subsection{Polynomial superpotential}

\begin{figure}[ht!]
\begin{center}
\begin{tabular}{ccc}
\hspace{-1.3cm}
\includegraphics[height=4. cm]{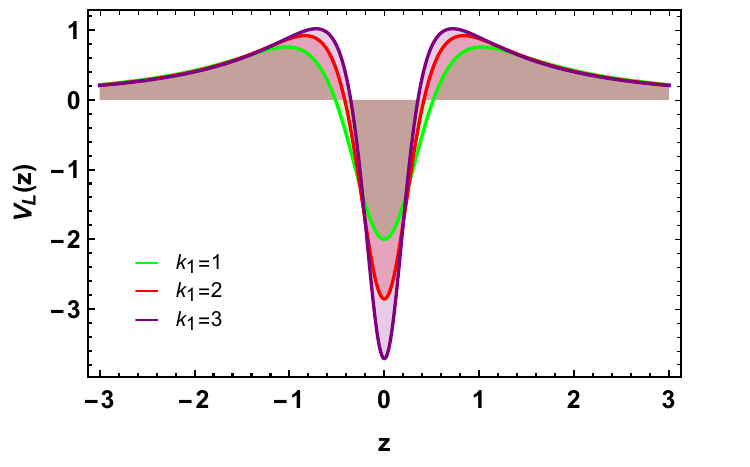} \hspace{1.2cm}
\includegraphics[height=4. cm]{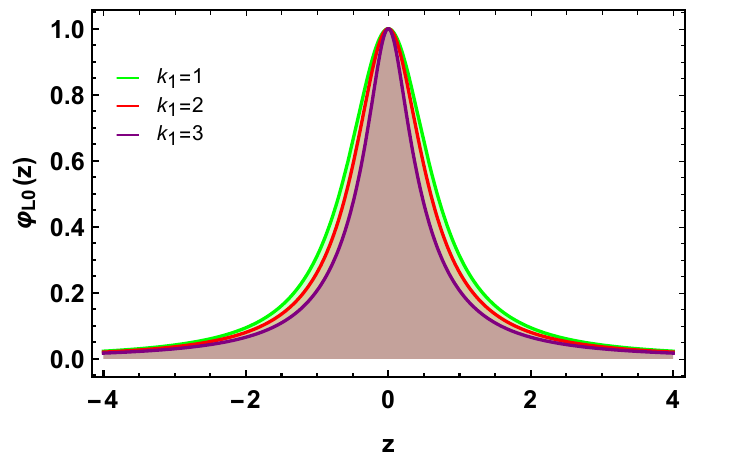}\\
(a)\hspace{6.7cm}(b)\\
\includegraphics[height=4.2cm]{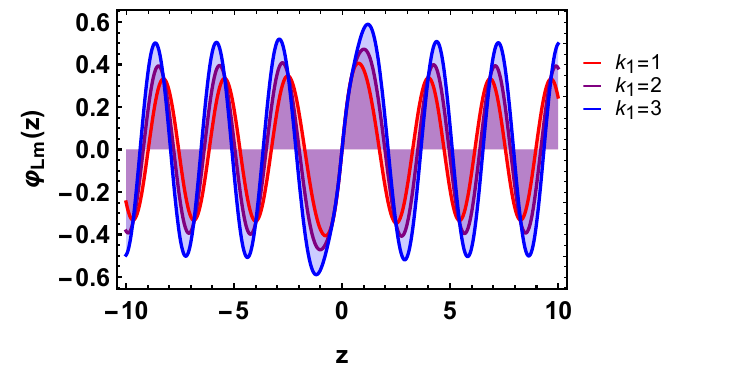}\hspace{-0.5cm}
\includegraphics[height=4.2cm]{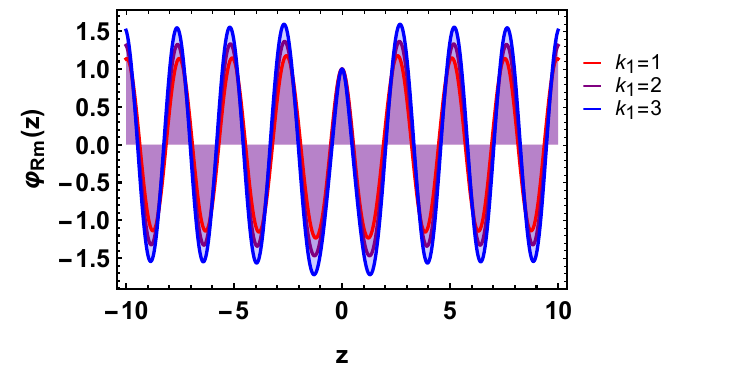}\\
(c)\hspace{6.7cm}(d) 
\end{tabular}
\end{center}
\vspace{-0.5cm}
\caption{ The case of the polynomial superpotential (\ref{Polynomialsuper}),   
with $a=\beta=k_2=1$, $\alpha=0,25$, and varying $k_1$.
(a) Effective potential. (b) Massless mode. (c) Massive odd mode ($m^2=5.079$). 
(d) Massive even mode ($m^2=6.772$).
\label{fig9}}
\end{figure}

\begin{figure}[ht!]
\begin{center}
\begin{tabular}{ccc}
\hspace{-1.3cm}
\includegraphics[height=4. cm]{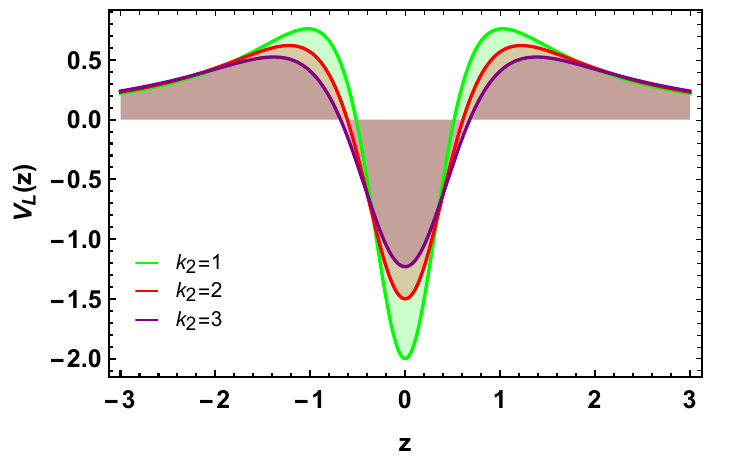} \hspace{1.2cm}
\includegraphics[height=4. cm]{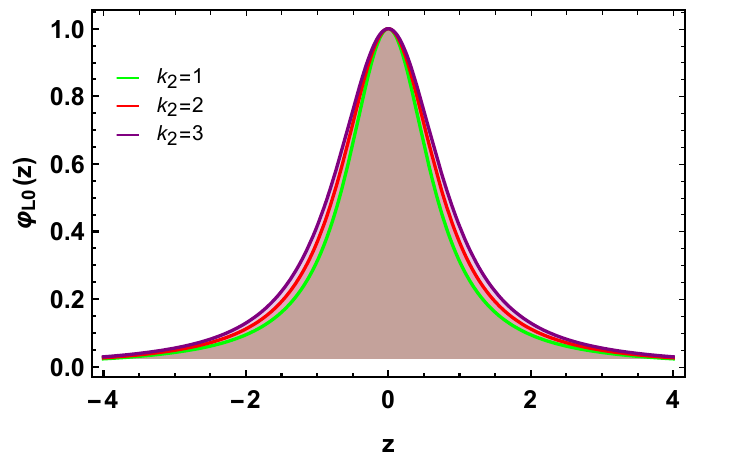}\\
(a)\hspace{6.7cm}(b)\\
\includegraphics[height=4.2cm]{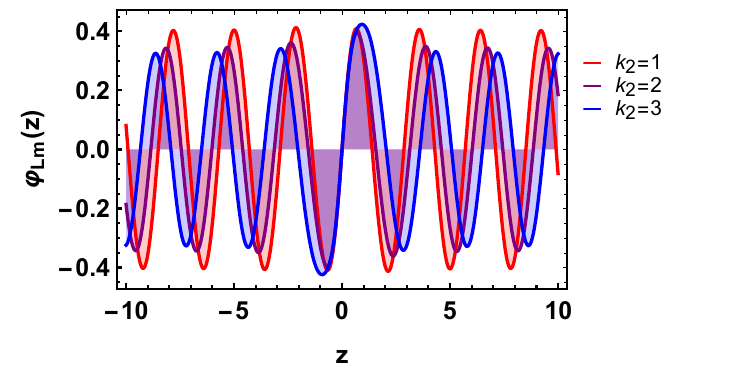}\hspace{-0.5cm}
\includegraphics[height=4.2cm]{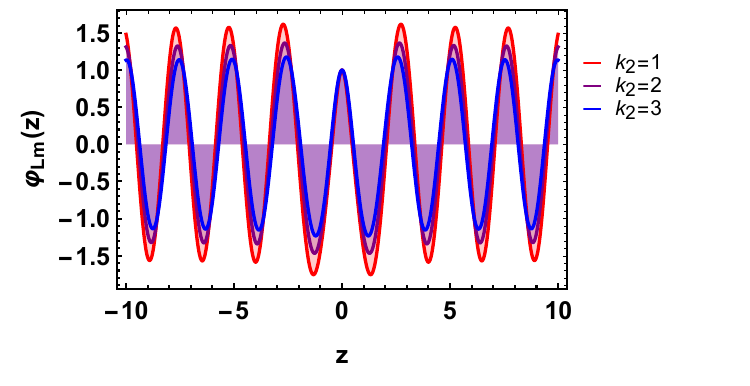}\\
(c)\hspace{6.7cm}(d) 
\end{tabular}
\end{center}
\vspace{-0.5cm}
\caption{  The case of the polynomial superpotential (\ref{Polynomialsuper}),   
 with $a=\beta=k_1=1$ $\alpha=0,25$, and varying $k_2$.
(a) Effective potential. (b) Massless mode. (c) Massive odd mode ($m^2=5.079$). 
(d) Massive even mode ($m^2=6.772$).
\label{fig10}}
\end{figure}

For     the polynomial superpotential 
(\ref{Polynomialsuper}),   we present the results  in Figs. 
\ref{fig9} and \ref{fig10}.  As we see, when we increase 
the nonmetricity parameter the effective potential becomes more confining, 
accentuating its well and potential barriers (see Fig. \ref{fig9} a). Moreover, 
it alters the massless fermionic modes, becoming more localized with the 
increase in the nonmetricity parameter  $k_1$ (see Fig. \ref{fig9} b). The 
massive modes also feel the effect of the nonmetricity term, therefore when 
we increase the value of $k_1$ we increase the amplitudes of the oscillations 
of the massive fermionic modes (see Fig. \ref{fig9} c and d). The opposite 
happens when we increase the momentum-energy parameter $k_2$, where the 
effective potential becomes less confining (see Fig. \ref{fig10} a), which 
leads to less localized massless fermionic modes (see Fig.  \ref{fig10} b) and 
massive fermionic modes that tend to reduce the amplitudes of their 
oscillations (see Fig. \ref{fig10} c and d).

\subsection{Linear superpotential}

\begin{figure}[ht!]
\begin{center}
\begin{tabular}{ccc}
\hspace{-1.3cm}
\includegraphics[height=4.cm]{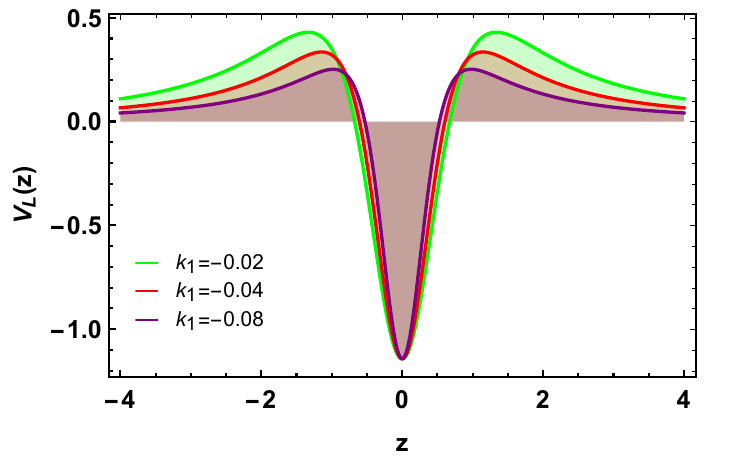}  \hspace{1.2cm}
\includegraphics[height=4.cm]{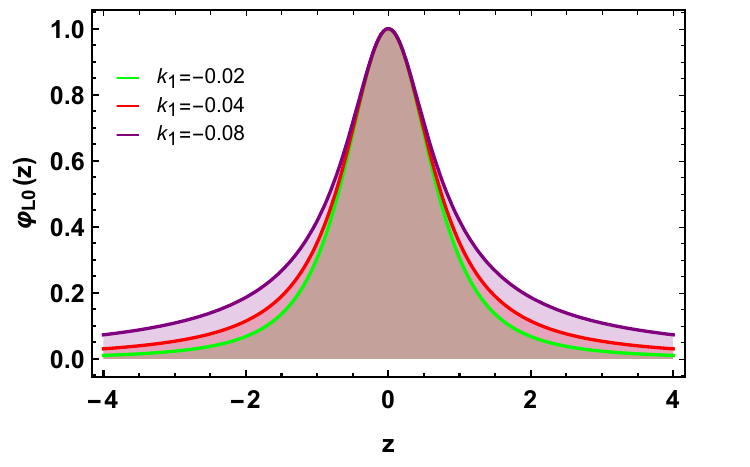}\\
(a)\hspace{6.7cm}(b)\\
\includegraphics[height=4.2cm]{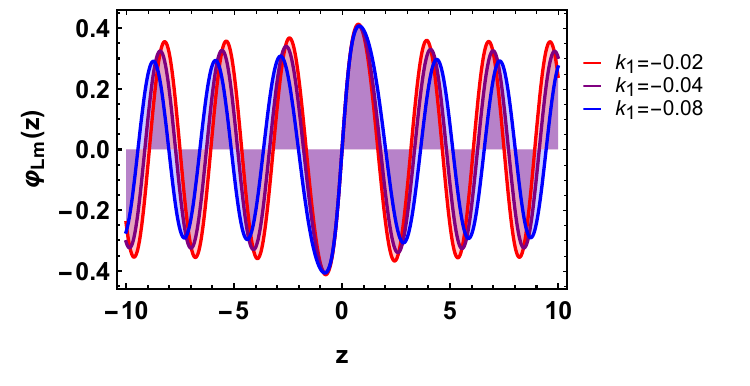}\hspace{-0.5cm}
\includegraphics[height=4.2cm]{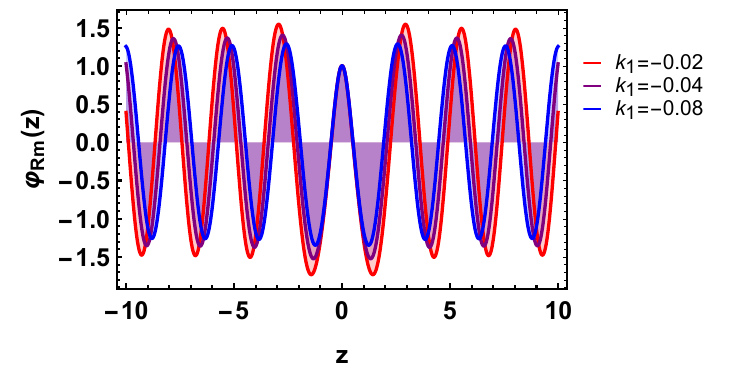}\\
(c)\hspace{6.7cm}(d) 
\end{tabular}
\end{center}
\vspace{-0.5cm}
\caption{  The case of  the linear superpotential (\ref{Linearsuper}), with 
$a=\beta=k_2=1$, $\alpha=0,25$, and varying $k_1$.
(a) Effective potential. (b) Massless mode. (c) Massive odd mode ($m^2=4.974$). 
(d) Massive even mode ($m^2=6.561$).
\label{fig11}}
\end{figure}

\begin{figure}[ht!]
\begin{center}
\begin{tabular}{ccc}
\hspace{-1.3cm}
\includegraphics[height=4.cm]{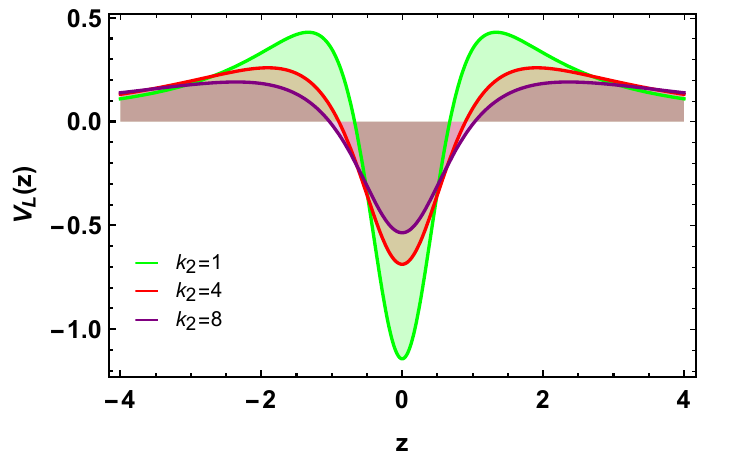}  \hspace{1.2cm}
\includegraphics[height=4.cm]{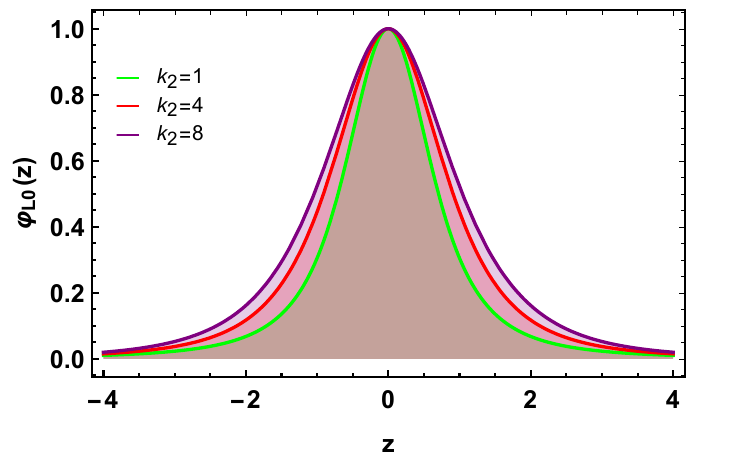}\\
(a)\hspace{6.7cm}(b)\\
\includegraphics[height=4.2cm]{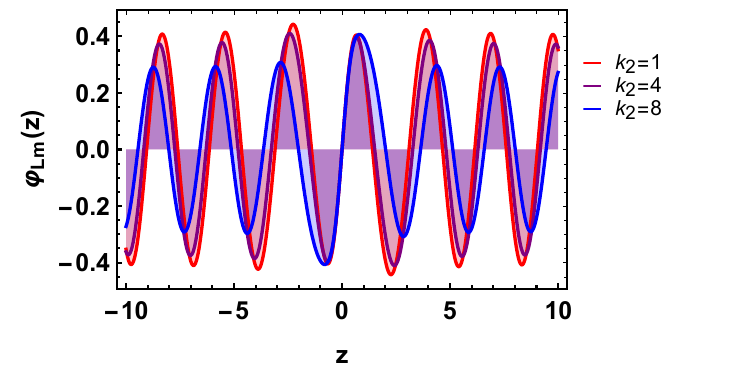}\hspace{-0.5cm}
\includegraphics[height=4.2cm]{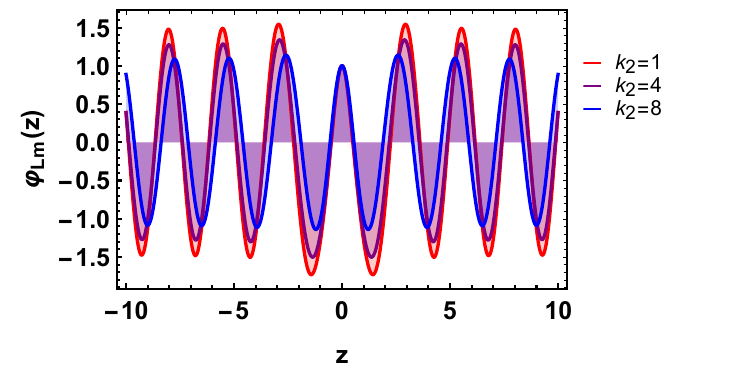}\\
(c)\hspace{6.7cm}(d) 
\end{tabular}
\end{center}
\vspace{-0.5cm}
\caption{  The case of  the linear superpotential (\ref{Linearsuper}),  with 
$a=\beta=1$, $k_1=-0.02$, $\alpha=0,25$, and varying $k_2$.
(a) Effective potential. (b) Massless mode. (c) Massive odd mode ($m^2=4.974$). 
(d) Massive even mode ($m^2=6.561$).
\label{fig12}}
\end{figure}
Finally, for  our third choice, namely the linear  superpotential 
(\ref{Linearsuper}),  the results  are depicted in Figs. \ref{fig11} 
and \ref{fig12}. It is easy to notice that when we reduce the nonmetricity 
parameter the effective potential becomes less confining (see Fig. \ref{fig11} 
a). The massless fermionic mode feels the changes suffered by the effective 
potential, becoming less localized with the decrease in the value of the 
nonmetricity parameter  $k_1$ (see Fig. \ref{fig11} b). Furthermore, the 
massive modes also feel the effect of the nonmetricity term, therefore when we 
decrease the value of $k_1$ we decrease the amplitudes of the oscillations of 
the massive fermionic modes (see Fig. \ref{fig11} c and d). These results are 
very similar to those obtained in the two previous superpotential choices.
Similarly, when we increase the value of the momentum-energy parameter $k_2$, 
we make the effective potential less confining (see Fig. \ref{fig12} a), which 
leads to less localized massless fermionic modes (see Fig .\ref{fig12} b). This 
behavior is felt by massive fermionic modes that tend to reduce the amplitudes 
of their oscillations (see Fig. \ref{fig12} c and d).

\section{Conclusions}
\label{s5}

In this work we have dealt with a five-dimensional braneworld generated by a 
single scalar field with cuscuton dynamics in nonmetricity-based modified 
gravity. We have considered the first-order formalism along with three 
well-known superpotentials to obtain a complete description of braneworld, 
choosing the modified gravity class $f(Q,\mathcal{T})=Q+k_1Q^n+k_2 
\mathcal{T}$, where the parameters $k_{1,2}$ and $n$ which are related to the 
effect of nonmetricity and trace of momentum-energy tensor, respectively.

As we showed, the addition of the cuscuton term  provides significant 
modifications to the structure of the brane.
In particular, we saw that the scalar field solutions have the form of a 
kink-like structure that can be intensified by varying the parameters $k_{1,2}$.
 The energy density is well localized, being able to be more localized when we 
vary the parameters $k_{1,2}$, and the same behavior holds for the cuscuton 
parameter  $\alpha$. 

Furthermore, for the location of the fermions for a minimal Yukawa-type 
coupling applying probabilistic measures we found sensitivity 
to changes in the parameters that control the gravitational theory. We arrived 
at a Schrödinger-like equation, which represents a supersymmetric equation of 
quantum mechanics, allowing for a normalizable massless mode. Our solutions for 
massless modes indicated strong brane localization only for left-chirality 
fermions. Both the effective potentials and the massless modes depend on the 
parameters $k_{1,2}$. It is noteworthy that the parameters $k_{1,2}$ can make 
zero-mode more (or less) localized. The same happens when we vary the cuscuton 
parameter  $\alpha$.  In turn, the massive modes present solutions similar to 
free waves, which indicates that these fermions probably escape the brane (the 
higher the mass eigenvalue, the greater the chances of these fermions escaping 
the brane). Finally, we found  that massive fermions have a greater sensitivity 
to gravitational changes in the core of the brane, where oscillations with more 
pronounced amplitudes are present.

As perspectives  for further papers, we can investigate the localization of 
abelian gauge fields in the braneworld scenario at hand. Moreover, it is 
possible to study how the graviton massives modes contribute to Newton's Law. 
Another issue that it is interesting to be investigated is the phase 
transitions of field configuration by means of the configurational entropy  
framework.  These studies will be performed in separate works.

\section*{Acknowledgments}  SHD would like to thank the 
partial support of project 20240220-SIP-IPN, Mexico and started this work on the research stay in China.  
ENS acknowledges the contribution  of 
the LISA CosWG, and of   COST 
Actions   CA21136 ``Addressing observational tensions in cosmology with 
systematics and fundamental physics (CosmoVerse)'', CA21106 ``COSMIC WISPers 
in the Dark Universe: Theory, astrophysics and experiments'', 
and CA23130 ``Bridging high and low energies in search of quantum gravity 
(BridgeQG)''.

\end{document}